\documentclass[pra,twocolumn,groupedaddress,showpacs,floatfix]{revtex4}
\usepackage{graphics}
\usepackage{graphicx}
\usepackage{bm}
\usepackage{amsmath}
\usepackage{amsfonts}
\usepackage{amssymb}
\usepackage{latexsym}
\usepackage{color}

\begin{document}

\title{A rate equation approach to cavity mediated laser cooling}
\author{Tony Blake,\footnote{Corresponding author: pytb@leeds.ac.uk} Andreas Kurcz, and Almut Beige}
\affiliation{The School of Physics and Astronomy, University of Leeds, Leeds, LS2 9JT, United Kingdom} 

\date{\today}

\begin{abstract}
The cooling rate for cavity mediated laser cooling scales as the Lamb-Dicke parameter $\eta$ squared. A proper analysis of the cooling process hence needs to take terms up to $\eta^2$ in the system dynamics into account. In this paper, we present such an analysis for a standard scenario of cavity mediated laser cooling with $\eta \ll 1$. Our results confirm that there are many similarities between ordinary and cavity mediated laser cooling. However, for a weakly confined particle inside a strongly coupled cavity, which is the most interesting case for the cooling of molecules, numerical results indicate that even more detailed calculations are needed to model the cooling process accurately.
\end{abstract}

\maketitle

\section{Introduction} \label{Intro}

First indications that cavity mediated laser cooling allows us to cool particles, like trapped atoms, ions and molecules, to much lower temperatures than other cooling techniques were found in Paris already in 1995 \cite{Karine,Grangier}. Systematic experimental studies of cavity mediated laser cooling have subsequently been reported by the groups of Rempe \cite{rempe00,pinkse2,Rempe,rempe09}, Vuleti\'c \cite{vuletic23,vuletic,vul,new}, and others \cite{kimble,chap}. Recent atom-cavity experiments access an even wider range of experimental parameters by replacing conventional high-finesse cavities \cite{Kim3,Kim4} through optical ring cavities \cite{Nagorny,Nagorny2} and tapered nanofibers \cite{Kim,Kim2} and by combining optical cavities with atom chip technology \cite{trupke,reichel}, atomic conveyer belts \cite{meschede0,meschede}, and ion traps \cite{Drewsen}. Moreover, Wickenbrock {\em et al.}~\cite{Renzoni} recently reported the observation of collective effects in the interaction of cold atoms with a lossy optical cavity. Motivated by these developments, this paper aims at increasing our understanding of cavity mediated laser cooling. 

Cavity-mediated laser cooling of free particles was first discussed in Refs.~\cite{lewen,lewen2}. Later, Ritsch and collaborators \cite{domokos2,peter2,ritsch97,ritsch98,peter3}, Vuleti{\'c} {\em et al.} \cite{vuletic10,vuletic3}, and others \cite{Murr,Murr2,Murr3,Robb} developed semiclassical theories to model cavity mediated cooling processes very efficiently. The analysis of cavity mediated laser cooling based on a master equation approach has been pioneered by Cirac {\em et al.}~\cite{Cirac2} in 1993. Subsequently, this approach has been used by many authors \cite{Cirac4,cool,morigi,morigi2,Tony}, since the precision of its calculations is easier to control than the precision of semiclassical calculations. Moreover, cavity mediated cooling is especially then of practical interest when it allows to cool particles to very low temperatures, where quantum effects dominate the time evolution of the system and semiclassical models no longer apply \cite{peter2}.

\begin{figure}[t]
\begin{minipage}{\columnwidth}
\begin{center}
\includegraphics[scale=0.6]{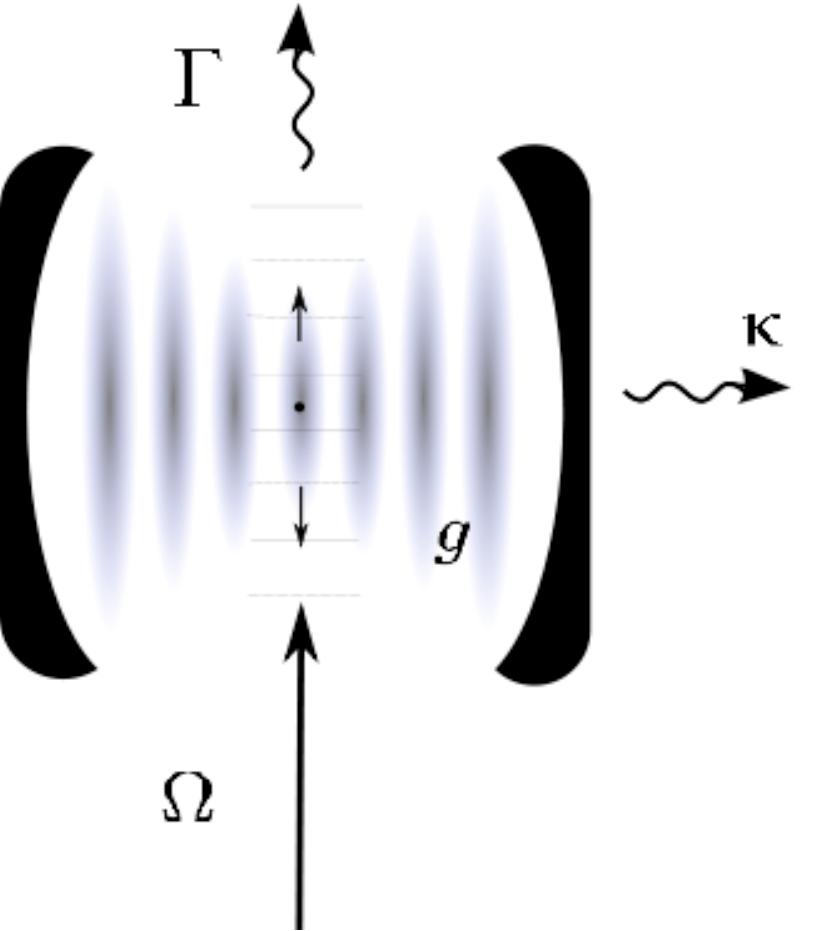}
\end{center}
\caption{Experimental setup of externally trapped particles inside an optical cavity with coupling constant $g$ and spontaneous decay rates $\kappa$ and $\Gamma$. The motion of the particles orthogonal to the cavity axis is confined by a harmonic trapping potential with phonon frequency $\nu$. Moreover, a cooling laser with Rabi frequency $\Omega$ is applied.} \label{setup}
\end{minipage}
\end{figure}

\begin{figure}[t]
\begin{center}
\includegraphics[scale=0.2]{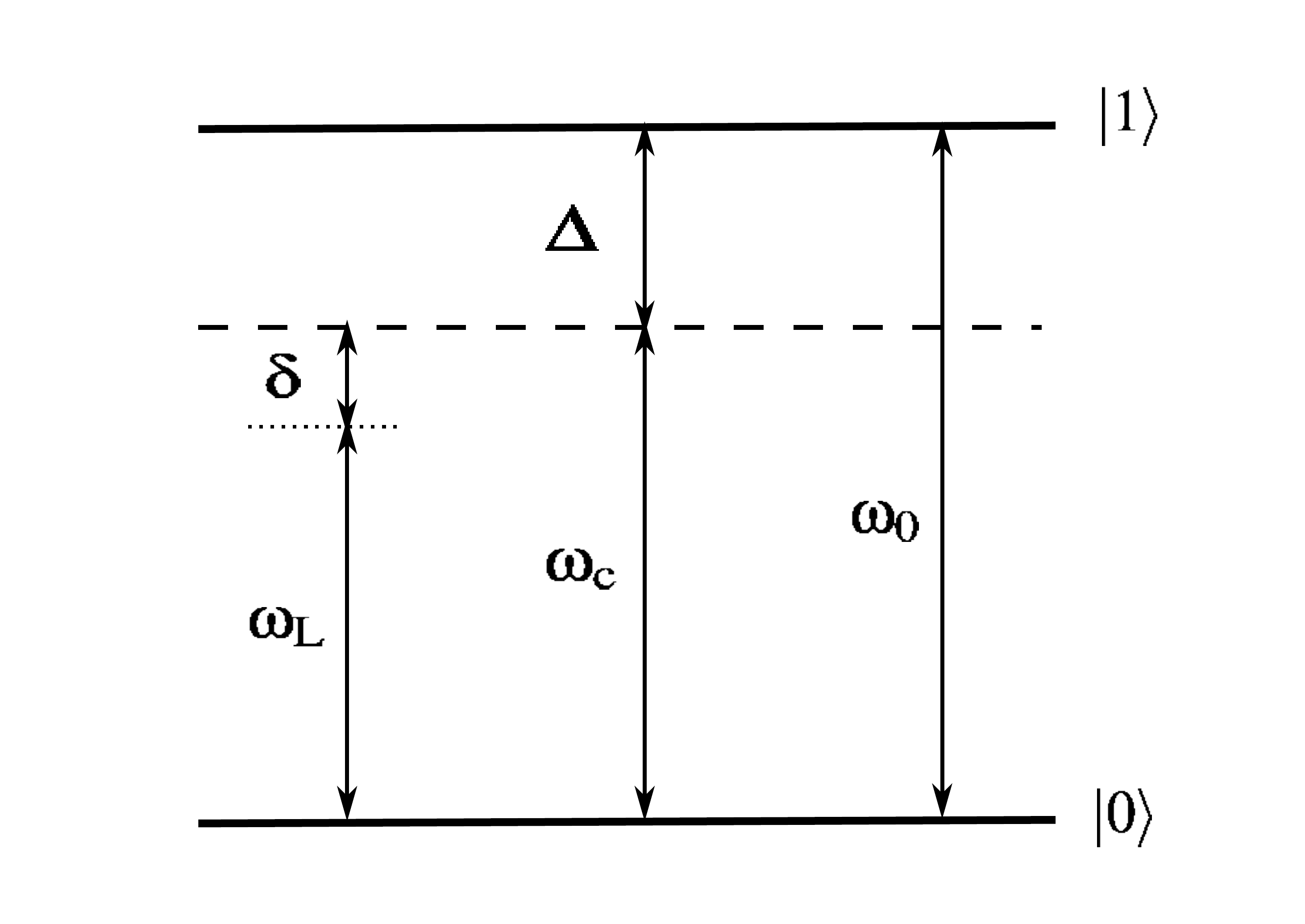}
\end{center}
\caption{Level configuration showing the ground $|0 \rangle$ and the excited state $|1 \rangle$ of the trapped particle. Here $\omega_{\rm L}$, $\omega_{\rm c}$, and $\omega_0$ are the frequency of the cooling laser, of the cavity field, and of the 0--1 transition of the particle, while $\delta$ and $\Delta$ denote detunings.} \label{energy1}
\end{figure}

The purpose of this paper is to provide a detailed analysis of the standard cavity cooling scenario illustrated in Figs.~\ref{setup} and \ref{energy1} which has already been studied by many authors \cite{domokos2,peter2,Murr2,Murr3,Cirac4,cool,morigi,morigi2,Tony,Andre,mo2,Lev,Kowa}. Analogously to ordinary laser cooling \cite{Stenholm4,Stenholm2,Tony2}, the effective cooling rate of cavity mediated laser cooling scales as the Lamb-Dicke parameter $\eta $ squared. A proper analysis of the cooling process using the above mentioned master equation approach \cite{Cirac2} hence needs to take terms up to order $\eta^2$ in the system dynamics into account. Doing so, the authors of Refs.~\cite{Cirac4,morigi,morigi2} derived an effective cooling equation of the form 
\begin{eqnarray} \label{strong24}
\dot m &=& - \eta^2 (A_- - A_+) \, m + \eta^2 A_+ 
\end{eqnarray}
which applies towards the end of the cooling process. Here $m$ is the mean phonon number and the $A_\pm$ denote transition rates. In the following we derive a closed set of 25 cooling equations which allow for a more detailed analysis of the cooling process itself and apply in the strong and in the weak confinement regime. Our calculations are analogous to the calculations presented in Refs.~\cite{Stenholm4,Stenholm2,Tony2} for ordinary laser cooling. When simplifying our rate equations via adiabatic eliminations, we obtain transitions rates $A_\pm$ that are consistent with the results reported in Refs.~\cite{Cirac4,morigi,morigi2}.

The reason that our calculations are nevertheless relatively straightforward is that we replace the phonon and the cavity photon annihilation operators $b$ and $c$ by two commuting annihilation operators $x$ and $y$. These commute with each other and describe bosonic particles that are neither phonons nor cavity photons. The corresponding system Hamiltonian no longer contains any displacement operators and provides a more natural description of the cavity-phonon system. Non-linear effects in the interaction between the vibrational states of the trapped particle and the field inside the optical cavity are now taken into account by a non-linear term of the form $x^\dagger x (y - y^\dagger)$. Using the master equation corresponding to this Hamiltonian, we then derive a closed set of 25 cooling equations. These are linear differential equations which describe the time evolution of $x$, $y$, and mixed operator expectation values and can be solved analytically as well as numerically. Applying the same methodology to laser cooling of a single trapped particle \cite{Tony2}, we recently obtained results which are consistent with previous results of other authors \cite{Stenholm4,Stenholm2}.  

The analytical calculations in this paper confirm that there are many similarities between ordinary and cavity mediated laser cooling \cite{Tony}. In the {\em strong} confinement regime, where the phonon frequency $\nu$ is much larger than the spontaneous cavity decay rate $\kappa$, we find that the optimal laser detuning $\delta_{\rm eff}$ is close to $\nu$ (sideband cooling). Different from this, one should choose $\delta_{\rm eff}$ close to ${1 \over 2} \kappa $ in the {\em weak} confinement regime with $\nu \ll \kappa$ in order to minimise the final temperature of the trapped particle. What limits the final phonon number are the counter-rotating terms in the particle-phonon interaction Hamiltonian $H_{\rm I}$ which can increase the energy of an open quantum system very rapidly \cite{SL,EC}.

In addition, we present detailed numerical studies. These take all available terms up to order $\eta^2$ in the rate equations into account, even the ones that appeared to be negligible during analytical calculations. Our numerical results are in general in very good agreement with our analytical results. However, for relatively small phonon frequencies $\nu$ and relatively large effective cavity coupling constants $g_{\rm eff}$, the stationary state phonon numbers predicted by both calculations can differ by more than one order of magnitude. This indicates that the analysis of the cooling process of a weakly confined particle inside a strongly coupled cavity needs to be more precise. Terms of order $\eta^3$ should be taken into account systematically when modelling the system dynamics with rate equations. Notice that the parameter regime where $g_{\rm eff}/\nu$ is relatively large is of special interest for the cooling of molecules. These usually have a relatively large atomic dipole moment but nevertheless cannot be trapped as easily as other particles.   
 
There are six sections in this paper. Section \ref{model} introduces the master equation for the atom-cavity-phonon system shown in Fig.~\ref{setup} and simplifies it via an adiabatic elimination of the excited electronic states of the trapped particle. Section \ref{standard} uses this master equation to derive a closed set of 25 cooling equations. These simplify respectively to a set of five effective cooling equations in the weak confinement regime and to a single effective cooling equation in the strong confinement regime. In Section \ref{stability} we show that the phonon coherences and the mean phonon number $m$ always reach their stationary state. Section \ref{progress} analyses the cooling process in more detail and calculates effective cooling rates and stationary state phonon numbers. Finally, we summarise our findings in Section \ref{conc}. Mathematical details are confined to Apps.~\ref{adel}--\ref{appC}.

\section{Theoretical model} \label{model}

The experimental setup considered in this paper is shown in Fig.~\ref{setup}. It contains a strongly confined particle inside an optical cavity. The aim of the cooling process is to minimise the number of phonons in the quantised motion of this particle in the direction of a cooling laser which enters the setup from the side. Cooling the motion of the particle in more than one direction would require additional cooling lasers. 

\subsection{The Hamiltonian}

The Hamiltonian of the atom-cavity-phonon system in Fig.~\ref{setup} is of the general form 
\begin{eqnarray} \label{1.23} 
H &=& H_{\rm par} + H_{\rm phn} + H_{\rm cav} + H_{\rm L} + H_{\rm par-cav} \, .
\end{eqnarray}
The first three terms are the free energy of the trapped particle, its quantised vibrational mode, and the quantised cavity field. Suppose, the particle is a two-level system with ground state $|0\rangle$ and excited state $|1\rangle$ and the energies $\hbar \omega_0$, $\hbar \nu$, and $\hbar \omega_{\rm c}$ are the energy of a single atomic excitation, a single phonon, and a single cavity photon, respectively, as illustrated in Fig.~\ref{energy1}. Then
\begin{eqnarray} \label{1.23b}
H_{\rm par} &=& \hbar \omega_0 \, \sigma^+ \sigma^- \, , \notag \\
H_{\rm phn} &=& \hbar \nu \, b^\dagger b \, , \notag \\
H_{\rm cav} &=& \hbar \omega_{\rm c} \, c^\dagger c \, ,
\end{eqnarray}
where the operators $\sigma^{-} \equiv |0\rangle\langle1|$ and $\sigma^{+}\equiv|1\rangle\langle0|$ are the atomic lowering and raising operator, $b$ is the phonon annihilation operator, and $c$ is the cavity photon annihilation operator with the bosonic commutator relation
\begin{eqnarray} \label{comms}
[b,b^{\dagger}] \,= \, [c,c^{\dagger}] \, = \, 1 \, . 
\end{eqnarray}
Let us now have a closer look at the two remaining terms $H_{\rm L}$ and $H_{\rm par-cav}$ in Eq.~(\ref{1.23}).
 
The role of the cooling laser is to establish a coupling between the electronic states $|0 \rangle$ and $|1 \rangle$ of the trapped particle and its quantised motion. Its Hamiltonian in the dipole approximation equals
\begin{eqnarray} \label{HL1}
H_{\rm L} &=& e\textbf{D} \cdot \textbf{E}_{\rm L} ({\bf x},t) \, ,
\end{eqnarray}
where $e$ is the charge of a single electron, ${\bf D}$ is the dipole moment of the particle,
\begin{eqnarray} \label{DDD}
\textbf{D}&=& \textbf{D}_{01} \, \sigma^{-}+\mbox{H.c.} \, ,
\end{eqnarray}
and $ \textbf{E}_{\rm L} (\textbf{x},t)$ denotes the electric field of the laser at position ${\bf x}$ at time $t$. Moreover, we have
\begin{eqnarray}
\textbf{E}_{\rm L} (\textbf{x},t)=\textbf{E}_0 \, {\rm e}^{{\rm i}(\textbf{k}_{\rm L} \cdot\textbf{x}-\omega_{\rm L} t)} + {\rm c.c.} 
\end{eqnarray} 
with $\textbf{E}_0$, $\textbf{k}_{\rm L}$, and $\omega_{\rm L}$ denoting amplitude, wave vector, and frequency of the cooling laser.  

The interaction Hamiltonian describing the coupling between the particle and the cavity in the dipole approximation is given by  
\begin{eqnarray}
H_{\rm par-cav} &=&  e \textbf{D} \cdot\textbf{E}_{\rm cav}(\textbf{x}) \, ,
\end{eqnarray}
where $\textbf{E}_{\rm cav}(\textbf{x})$ is the observable for the quantised electric field inside the resonator at the position of the particle. Denoting the corresponding coupling constant as $g$, the above Hamiltonian becomes 
\begin{eqnarray}\label{2.10}
H_{\rm par-cav} &=& \hbar g (\sigma^{-}+\sigma^{+}) \, c+ {\rm H.c.} 
\end{eqnarray} 
This Hamiltonian describes the possible exchange of energy between atomic states and the cavity.

\subsection{Displacement operator}

The relevant vibrational mode of the trapped particle is its center of mass motion in the laser direction. Considering this motion as quantised with the phonon annihilation operator $b$ from above yields
\begin{eqnarray} \label{2.21}
\textbf{k}_{\rm L} \cdot \textbf{x} &=& \eta (b + b^\dagger) \, , 
\end{eqnarray}
where the Lamb-Dicke parameter $\eta $ is a measure for the steepness of the effective trapping potential seen by the particle \cite{Stenholm2}. Substituting Eqs.~(\ref{DDD})--(\ref{2.21}) into Eq.~(\ref{HL1}), we find that the laser Hamiltonian is a function of the  particle displacement operator \cite{Knight}
\begin{eqnarray}\label{2.22}
D ({\rm i}\eta) &\equiv & {\rm e}^{- {\rm i} \eta (b+b^\dagger)} \, .
\end{eqnarray}
This operator is a unitary operator,  
\begin{eqnarray} \label{unitary}
D({\rm i} \eta)^\dagger D({\rm i} \eta) = D({\rm i} \eta) D({\rm i} \eta)^\dagger &=& 1 \, , 
\end{eqnarray}
with
\begin{eqnarray}\label{2.2220}
D ({\rm i}\eta) \, b \, D ({\rm i}\eta)^\dagger &=& b+ {\rm i} \eta \, , \nonumber \\
D ({\rm i}\eta)^\dagger \, b \, D({\rm i}\eta) &=& b - {\rm i} \eta \, .
\end{eqnarray}
Using this operator, $H_{\rm L}$ can be written as
\begin{eqnarray}\label{2.4}
H_{\rm L} &=& e \left[ \textbf{D}_{01} \, \sigma^{-} + {\rm H.c.} \right] \cdot \textbf{E}_0^* \, D({\rm i}\eta) \, {\rm e}^{ {\rm i} \omega_{\rm L} t} + {\rm H.c.} ~~~
\end{eqnarray}
The cooling laser indeed couples the vibrational and the electronic states of the trapped particle.

\subsection{Effective interaction Hamiltonian} \label{IP}

Let us continue by introducing an interaction picture, in which the Hamiltonian $H$ in Eq.~(\ref{1.23}) becomes time independent. To do so, we choose
\begin{eqnarray} \label{intpic}
H_0 &=&  \hbar\omega_{\rm L} \, \sigma^{+}\sigma^{-} + \hbar \omega_{\rm L} \, c^{\dagger}c \, .
\end{eqnarray}
Neglecting relatively fast oscillating terms, i.e.~terms which oscillate  with $2 \omega_{\rm L}$, as part of the usual rotating wave approximation and using the same notation as in Fig.~\ref{energy1}, the interaction Hamiltonian $H_{\rm I}$,
\begin{eqnarray} \label{trafo}
H_{\rm I}=U^{\dagger}_0(t,0) \, (H-H_0) \, U_0(t,0) \, ,
\end{eqnarray}
becomes 
\begin{eqnarray}\label{2.8}
H_{\rm I} &=& {1 \over 2} \hbar \Omega \, D({\rm i}\eta) \sigma^- + \hbar g \, \sigma^- c^+ +\mbox{H.c.}  \notag \\
&& + \hbar\left(\Delta+\delta \right)\sigma^+\sigma^- +\hbar\nu \, b^{\dagger}b + \hbar \delta \, c^{\dag}c \, . 
\end{eqnarray}
Here $\Delta $ and $\Delta + \delta$ denote the detuning of the cavity and of the laser with respect to the 0--1 transition of the trapped particle, respectively. 

In the following we assume that $|\Delta|$ is much larger than all other system parameters,  
\begin{eqnarray} \label{condi0}
|\Delta | &\gg& \Omega \, , ~ |\delta | \, ,~ \nu \, , ~ g \, , ~ \Gamma \, , ~ \kappa \, .
\end{eqnarray}
This condition allows us to eliminate the electronic states of the trapped particle adiabatically from the system dynamics. 
Doing so and proceeding as in App.~\ref{adel}, we obtain the effective interaction Hamiltonian
\begin{eqnarray} \label{3.7}
H_{\rm I} &=& \hbar g_{\rm eff} \, D({\rm i}\eta) c + {\rm H.c.} + \hbar \nu \, b^{\dagger}b + {\hbar \delta_{\rm eff}} \, c^{\dagger}c ~~
\end{eqnarray}
with $g_{\rm eff}$ and $\delta_{\rm eff}$ defined as 
\begin{eqnarray} \label{3.8} 
g_{\rm eff} &\equiv & - \frac{ g \Omega}{2\Delta} \quad \mbox{and} \quad \delta_{\rm eff} \equiv \delta - \frac{g^2}{\Delta} \, .
\end{eqnarray}
The interaction Hamiltonian $H_{\rm I}$ in Eq.~(\ref{3.7}) holds up to first order in $1/\Delta$. It no longer contains any atomic operators and describes instead a direct coupling between cavity photons and phonons. 

\subsection{Master equation}

After the adiabatic elimination of the electronic states of the trapped particle, the only relevant decay channel in the system is the leakage of photons through the cavity mirrors. To take this into account, we describe the cooling process in the following by the master equation 
\begin{eqnarray}\label{3.1}
\dot{\rho}= - {{\rm i} \over \hbar} \left[H_{\rm I},\rho\right] - {1 \over 2} \kappa \left( c^\dagger c \rho + \rho c^\dagger c \right) + \kappa \, c \rho c^\dagger  
\end{eqnarray}
with $H_{\rm I}$ as in Eq.~(\ref{3.7}), where $\kappa$ denotes the spontaneous decay rate for  a single photon inside the cavity. 

\section{Cooling equations} \label{standard}

In the following, we use the above master equation to derive linear differential equations for expectation values, so-called rate or cooling equations. Obtaining a closed set of rate equations is not straightforward due to the presence of the displacement operator $D$ in Eq.~(\ref{2.22}). To significantly reduce the number of rate equations which have to be taken into account in the following calculations, we first introduce two new operators $x$ and $y$ which replace the phonon and the cavity photon annihilation operators $b$ and $c$ by two new bosonic operators $x$ and $y$. These commute with each other and provide a more natural description of the cavity-phonon system.  

\subsection{Transformation of the Hamiltonian}

To simplify the Hamiltonian $H_{\rm I}$ in Eq.~(\ref{3.7}), we now proceed analogously to Ref.~\cite{Tony2} and define  
\begin{eqnarray}\label{25}
x &\equiv& D({\rm i} \eta) \, c \, . 
\end{eqnarray} 
This operator annihilates a cavity photon while simultaneously giving a kick to the trapped particle. Since the displacement operator $D({\rm i} \eta)$ is a unitary operator (c.f.~Eq.~(\ref{unitary})) one can easily check that $x$ fulfils the bosonic commutator relation 
\begin{eqnarray}\label{26}
\left[ x, x^\dagger \right] &\equiv& 1 \, . 
\end{eqnarray} 
This means, the particles created by $x^\dagger$ when applied to the vacuum are bosons. They are cavity photons whose creation is always accompanied by a displacement of the particle. Substituting Eq.~(\ref{25}) into Eq.~(\ref{3.7}), $H_{\rm I}$ becomes
\begin{eqnarray} \label{27}
H_{\rm I} &=& \hbar g_{\rm eff} \, x + {\rm H.c.} + {\hbar \delta_{\rm eff}} \, x^{\dagger} x + \hbar \nu \, b^{\dagger}b \, . ~~
\end{eqnarray}
In the following, we list commutator relations which can be derived using Eqs.~(\ref{comms}), (\ref{unitary}), and (\ref{2.2220}),
\begin{eqnarray}\label{28}
\left[ x, b \right] = - \left[ x, b^\dagger \right] &=& {\rm i} \eta \, x \, , \nonumber \\
\left[ x^{\dagger}, b \right] = - \left[ x^\dagger, b^\dagger \right] &=& - {\rm i} \eta \, x^\dagger \, .
\end{eqnarray} 
These can then be used to moreover show that
\begin{eqnarray}\label{29}
&& \left[ x, b^\dagger b \right] = - {\rm i} \eta \, x (b - b^\dagger) - \eta^2 \, x \, , \nonumber \\
&& \left[ x^\dagger, b^\dagger b \right] = {\rm i} \eta (b - b^\dagger) x^\dagger + \eta^2 \, x^\dagger \, , \nonumber \\
&& \left[ x^\dagger x, b \right] = \left[ x^\dagger x, b^\dagger \right]  \, = \, \left[ x^\dagger x, b^\dagger b \right] = 0 \, .  
\end{eqnarray} 
Unfortunately, the operators $x$ and $b$ do not commute with each other.

To assure that it is nevertheless possible to analyse the cooling process using only a relatively small number of cooling equations, we now introduce another operator $y$ as 
\begin{eqnarray} \label{defy}
y &\equiv & b - {\rm i} \eta \, c^\dagger c \, .
\end{eqnarray}
This operator annihilates phonons while simultaneously affecting the state of the cavity field. Using Eq.~(\ref{29}), one can show that $y$ too obeys a bosonic commutator relation, 
\begin{eqnarray}\label{26c}
\left[ y, y^\dagger \right] &=& 1 \, .
\end{eqnarray} 
Using the above commutator relations, one can moreover show that $x$ and $y$ commute with each other,
\begin{eqnarray}\label{33}
\left[ x, y \right] = \left[ x^\dagger, y \right] &=& 0 \, .
\end{eqnarray} 
Notice that the above transformation of $b$ and $c$ in Eqs.~(\ref{25}) and (\ref{defy}) are unitary operator transformations which leave the total Hilbert space of the cavity-phonon system invariant. Indeed one can show that \cite{Hadamard}
\begin{eqnarray} \label{uuu}
U &\equiv & \exp \left[ {\rm i} \eta c^\dagger c \left( b + b^\dagger \right) \right] \, .
\end{eqnarray} 
yields $x$ when defining $x$ as $x = U \, c \, U^\dagger$ and $y$ when defining $y$ as $y = U \, b \, U^\dagger$. 

Using the $x$ and the $y$ operator, the interaction Hamiltonian $H_{\rm I}$ in Eq.~(\ref{27}) can now be written as 
\begin{eqnarray} \label{35}
H_{\rm I} &=& \hbar g_{\rm eff} \, x + {\rm H.c.} + {\hbar \delta_{\rm eff}} \, x^{\dagger} x 
+ \hbar \eta^2 \nu \, x^\dagger x x^\dagger x \nonumber \\
&& - {\rm i} \hbar \eta \nu \, x^\dagger x (y -y^\dagger) + \hbar \nu \, y^{\dagger} y \, .
\end{eqnarray}
This Hamiltonian is exact, since the exponential terms in the original Hamiltonian $H_{\rm I}$ in Eq.~(\ref{2.8}) have been removed via a basis transformation and not via an approximation.  

\subsection{Time evolution of expectation values} \label{EV}

In the remainder of this section, we use the interaction Hamiltonian $H_{\rm I}$ to obtain a closed set of cooling equations, including one for the time evolution of the mean phonon number $m$. The time derivative of the expectation value of an arbitrary operator  $A$, which is time-independent operator in the relevant interaction picture, equals 
\begin{eqnarray}
\langle \dot A \rangle &=& \mbox{Tr} (A \dot{\rho}) \, .
\end{eqnarray}
When combining this result with Eq.~(\ref{3.1}), we find that $\langle A \rangle$ evolves according to
\begin{eqnarray} \label{dotA0}
\langle \dot A \rangle &=& -{{\rm i} \over \hbar} \, \left\langle\left[A,H_{\rm I} \right]\right\rangle - {1 \over 2} \kappa \, \langle A x^\dagger x + x^\dagger x A \rangle \nonumber \\
&& + \kappa \, \langle x^\dagger D({\rm i} \eta) A D({\rm i} \eta)^\dagger x \rangle 
\end{eqnarray}
with respect to the interaction picture which we introduced earlier in Section \ref{IP}. 

In this paper we are especially interested in the time evolution of the mean phonon number $m$ which is given by the expectation value
\begin{eqnarray} \label{ns}
m &\equiv & \langle b^{\dagger} b \rangle \, .
\end{eqnarray} 
Combining this equation with the definitions of $x$ and $y$ in Eqs.~(\ref{25}) and (\ref{defy}), we find that   
\begin{eqnarray} \label{mfinaldot}
m &\equiv & n_2 - \eta \, k_{12} + \eta^2 \, n_3 \, ,
\end{eqnarray} 
if we define 
\begin{eqnarray} \label{coherences}
&& \hspace*{-0.4cm} n_2 \equiv \langle y^{\dagger} y \rangle \, , ~~  
n_3 \equiv \langle x^{\dagger} x x^{\dagger} x \rangle \, , \nonumber \\
&& \hspace*{-0.4cm} k_{12} \equiv {\rm i}  \, \langle x^\dagger x (y - y^\dagger ) \rangle \, . ~~
\end{eqnarray}
This means, $m$ and $n_2$ are the same in zeroth order in $\eta$. In order to get a closed set of cooling equations, we need to consider in addition the variables
\begin{eqnarray} \label{coherences2}
&& \hspace*{-0.4cm} n_1 \equiv \langle x^{\dagger} x \rangle \, , ~~  
k_7 \equiv \langle y + y^\dagger \rangle \, , \nonumber \\
&& \hspace*{-0.4cm} k_8 \equiv {\rm i} \, \langle y - y^\dagger \rangle \, , ~~ 
k_9 \equiv  \langle y^2 + y^{\dagger 2} \rangle \, , \nonumber \\
&& \hspace*{-0.4cm} k_{10} \equiv  {\rm i}  \,  \langle y^2 - y^{\dagger 2} \rangle \, , ~~
k_{11} \equiv \langle x^\dagger x (y + y^\dagger ) \rangle ~~
\end{eqnarray}
and 16 other expectation values which we define in App.~\ref{appA}. These are not listed here, since they appear only in the appendices of this paper. 

For example, applying Eq.~(\ref{dotA0}) to the above introduced $y$ operator expectation values, we find that their time derivatives are without any approximations given by
\begin{eqnarray} \label{4888}
\dot n_2 &=& \eta \nu \, k_{11} - \eta \kappa \, k_{12} + \eta^2 \kappa \, n_1 \, , \notag \\
\dot k_7 &=& 2 \eta \nu \, n_1 - \nu \, k_8 \, , \nonumber \\ 
\dot k_8 &=& \nu \, k_7 - 2 \eta \kappa \, n_1 \, , \nonumber \\
\dot k_9 &=& - 2 \nu \, k_{10} + 2 \eta \nu \, k_{11} + 2 \eta \kappa \, k_{12} - 2 \eta^2 \kappa \, n_1 \, , ~~ \nonumber \\
\dot k_{10} &=& 2 \nu \, k_9 + 2 \eta \nu \, k_{12} - 2 \eta \kappa \, k_{11} \, . 
\end{eqnarray}
These five differential equations depend only on the $y$ operator expectation values themselves as well as on $n_1$, $k_{11}$, and $k_{12}$. The time derivatives of all other relevant expectation values can be found in App.~\ref{appB}.

\begin{figure}[t]
\begin{minipage}{\columnwidth}
\includegraphics[scale=0.9]{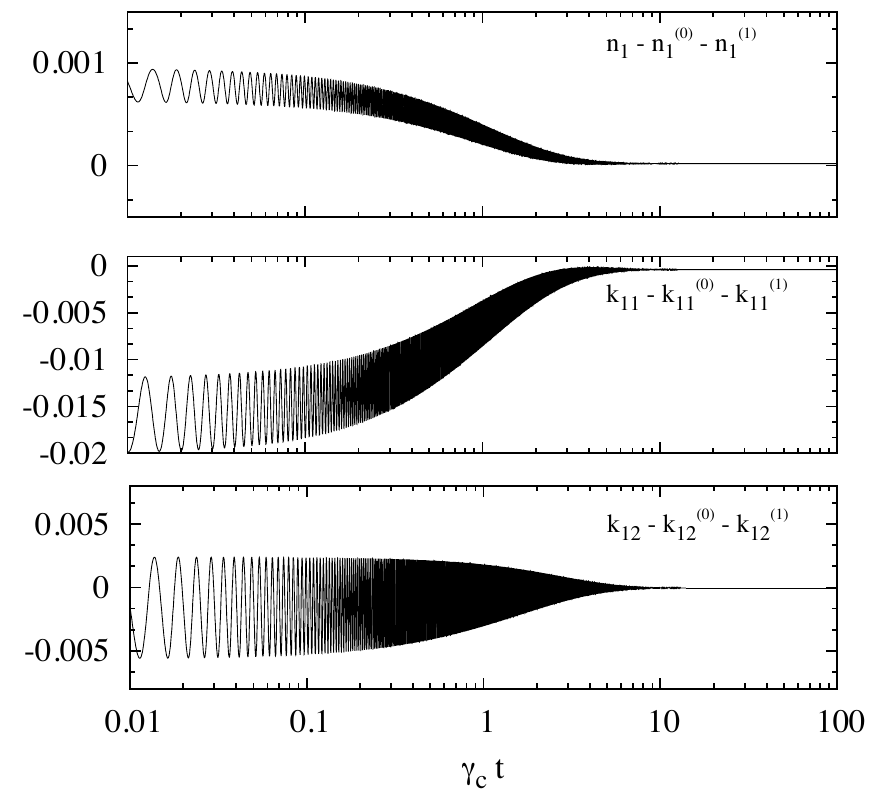}
\caption{Difference $n_1^{(0)} + n_1^{(1)}$, $k_{11}^{(0)} + k_{11}^{(1)}$, and $k_{12}^{(0)}+ k_{12}^{(1)}$ obtained from Eqs.~(\ref{zeroth}), (\ref{k11120}), (\ref{zeroth2}), and (\ref{k11121}) and $n_1$, $k_{11}$, and $k_{12}$ obtained from a numerical solution of the 25 cooling equations which can be found in Section \ref{effeq} and in App.~\ref{appB}. Here we have $\eta=0.1$, $\nu = 0.1 \, \kappa$, $\delta_{\rm eff}  = 0.5 \, \kappa$, and $g_{\rm eff} = 0.1 \, \kappa$ which are typical experimental parameters for a {\em weakly} coupled cavity in the {\em weak} confinement regime.} \label{pic}
\end{minipage}
\end{figure}

\subsection{Weak confinement regime} \label{effeq}

Let us first have a closer look at the case, where the trapped particle experiences a relatively weak trapping potential. In this subsection, we hence assume that the phonon frequency $\nu$ is much smaller than the spontaneous cavity decay rate $\kappa$, while the Lamb-Dicke parameter $\eta$ is much smaller than one,
\begin{eqnarray} \label{oups}
\nu \ll \kappa ~~ {\rm and} ~~ \eta \ll 1 \, .
\end{eqnarray}
When this applies, the $y$-operator expectation values evolve on a much slower time scale than all other expectation values. In the following, we take advantage of this time scale separation and eliminate all relevant $x$ and mixed operator expectation values adiabatically from the time evolution of the cavity-phonon system. The result of this calculation which can be found in App.~\ref{appB} are approximate solutions for $n_1$, $k_{11}$, and $k_{12}$ up to first order in $\eta$. 

\begin{figure}[t]
\begin{minipage}{\columnwidth}
\includegraphics[scale=0.9]{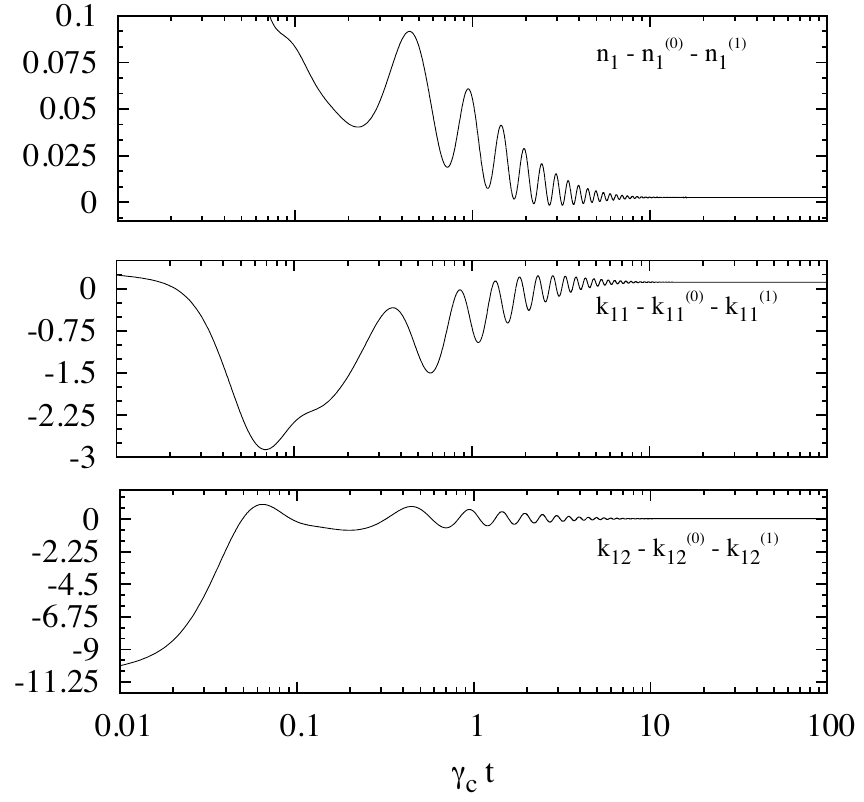}
\caption{Difference between the analytical and the numerical solutions for $n_1$, $k_{11}$, and $k_{12}$ as in Fig.~\ref{pic} but for $\eta=0.1$, $\nu = 0.1 \, \kappa$, $\delta_{\rm eff}  = 0.5 \, \kappa$, and $g_{\rm eff} = \kappa$ which are typical experimental parameters for a {\em strongly} coupled cavity in the {\em weak} confinement regime.} \label{pic2}
\end{minipage}
\end{figure}

Figs.~\ref{pic} compares the analytical expressions for $n_1$, $k_{11}$, and $k_{12}$ which we obtained in App.~\ref{appB} with the results of a numerical solution of the full set of 25 rate equations. For a weakly coupled optical cavity with $g_{\rm eff} \ll \kappa$, the numerical results differ indeed only very little from the results in Eqs.~(\ref{zeroth}), (\ref{k11120}), (\ref{zeroth2}), and (\ref{k11121}). The effective rate equations obtained in this subsection apply in this case after a short transition time of the order of $1/\kappa$. Fig.~\ref{pic2} makes a similar comparison for the case of a relatively strongly-coupled optical cavity with $g_{\rm eff} = \kappa$. In this case, there is less agreement between numerical and analytical results and the rate equations derived in this section apply really well only towards the end of the cooling process. Although we do not illustrate this here explicitly, let us mention that even less agreement is found when $g_{\rm eff} \gg \kappa$.

When substituting Eqs.~(\ref{zeroth}), (\ref{k11120}), (\ref{zeroth2}), and (\ref{k11121}) into Eq.~(\ref{4888}), we are left with a closed set of five effective cooling equations which hold up to order $\eta^2$, ie. 
\begin{eqnarray} \label{eff}
\big( \dot n_2 , \dot k_7 , \dot k_8 , \dot k_9 , \dot k_{10} \big)^{\rm T} 
&=& M \left( n_2 , k_7 , k_8 , k_9 , k_{10} \right)^{\rm T}  \nonumber \\
&& + \left( \beta_1 , \beta_2 , \beta_3 , \beta_4 , \beta_5 \right)^{\rm T} ~~
\end{eqnarray}
with
\begin{eqnarray} \label{eff3}
M &=& \left( \begin{array}{ccccc} \alpha_{11}^{(2)} & \alpha_{12}^{(1)} & \alpha_{13}^{(1)} & \alpha_{14}^{(2)} & 0 \\ 
0 & 0 & - \nu & 0 & 0 \\
0 & \nu & \alpha_{33}^{(2)} & 0 & 0 \\
\alpha_{41}^{(2)} & \alpha_{42}^{(1)} & \alpha_{43}^{(1)} &\alpha_{44}^{(2)} & - 2 \nu \\
0 & \alpha_{52}^{(1)} & \alpha_{53}^{(1)} & 2 \nu & \alpha_{55}^{(2)} \end{array} \right) .
\end{eqnarray}
Each superscript indicates the scaling of the respective matrix element of $M$ with respect to $\eta$. Taking Eq.~(\ref{oups}) into account, we find that the $\alpha_{ij}^{(1)}$ of $M$ are to a very good approximation given by
\begin{eqnarray} \label{eff4}
&& \alpha_{12}^{(1)} = - {8 \eta \nu g_{\rm eff}^2 (\kappa^2 - 4 \delta_{\rm eff}^2 ) \over (\kappa^2 + 4 \delta_{\rm eff}^2 )^2} \, , ~~
\alpha_{13}^{(1)} = - {4 \eta \kappa g_{\rm eff}^2 \over \kappa^2 + 4 \delta_{\rm eff}^2} \, , ~~~~ \nonumber \\  
&& \alpha_{42}^{(1)} = {32 \eta \kappa^2 \nu g_{\rm eff}^2  \over (\kappa^2 + 4 \delta_{\rm eff}^2)^2} \, , ~~  
\alpha_{43}^{(1)} = {8 \eta \kappa g_{\rm eff}^2 \over \kappa^2 + 4 \delta_{\rm eff}^2} \, , \nonumber \\
&& \alpha_{52}^{(1)} = - \alpha_{43}^{(1)} \, , ~~ \alpha_{53}^{(1)} = \alpha_{42}^{(1)} \, ,
\end{eqnarray}
while 
\begin{eqnarray} \label{eff6}
&& \hspace*{-0.9cm} \alpha_{11}^{(2)} = \alpha_{33}^{(2)} = \alpha_{44}^{(2)} = \alpha_{55}^{(2)} = - {64 \eta^2 \kappa \nu \delta_{\rm eff} g_{\rm eff}^2 \over (\kappa^2 + 4 \delta_{\rm eff}^2)^2} \, , ~~ \nonumber \\
&& \hspace*{-0.9cm} \alpha_{14}^{(2)} = {32 \eta^2 \kappa \nu \delta_{\rm eff} g_{\rm eff}^2 \over (\kappa^2 + 4 \delta_{\rm eff}^2)^2} \, , ~~
\alpha_{41}^{(2)} = {128 \eta^2 \kappa \nu \delta_{\rm eff} g_{\rm eff}^2 \over (\kappa^2 + 4 \delta_{\rm eff}^2)^2} \, . 
\end{eqnarray}
Moreover, one can show that $\beta_1$ equals, up to second order in $\eta$,  
\begin{eqnarray} \label{beta1}
\beta_1 &=& {4 \eta^2 \kappa g^2_{\rm eff} \over (\kappa^2 + 4 \delta_{\rm eff}^2)^3} \left[ (\kappa^2 + 4 \delta_{\rm eff}^2) \left(\kappa^2 + 4\delta^2_{\rm eff} - 8 \delta_{\rm eff} \nu \right) \right. \nonumber \\
&& \left. + 8 g^2_{\rm eff} \left(3 \kappa^2 - 4\delta^2_{\rm eff}\right) \right] \, , ~~
\end{eqnarray}
while $\beta_2$ to $\beta_5$ are in first order in $\eta$ given by
\begin{eqnarray} \label{betas}
&& \beta_2 = {8 \eta \nu g_{\rm eff}^2 \over \kappa^2 + 4 \delta_{\rm eff}^2} \, , ~~
\beta_3 = - {8 \eta \kappa g_{\rm eff}^2 \over \kappa^2 + 4 \delta_{\rm eff}^2} \, , \notag \\
&& \beta_4 = \beta_5 = 0 \, .
\end{eqnarray}
We now have a closed set of five differential equations which can be used to analyse the time evolution of the $y$ operator expectation values in the weak confinement regime analytically and numerically. 

\begin{figure}[t]
\begin{minipage}{\columnwidth}
\includegraphics[scale=0.9]{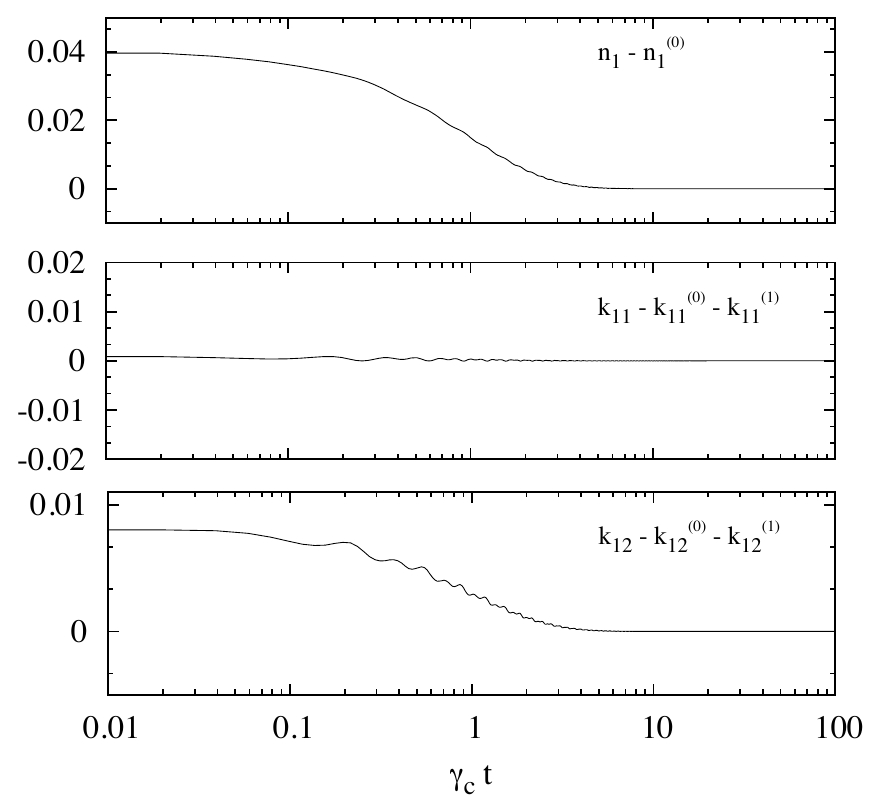}
\caption{Difference between the analytical and the numerical solutions for $n_1$, $k_{11}$, and $k_{12}$ as in Fig.~\ref{pic} but for $\eta=0.1$, $\nu = 10 \, \kappa$, $\delta_{\rm eff}  = \nu$, and $g_{\rm eff} = 0.1 \, \kappa$ which are typical experimental parameters for a {\em weakly} coupled cavity in the {\em strong} confinement regime.} \label{pic3}
\end{minipage}
\end{figure}

\subsection{Strong confinement regime} \label{effeq2}

In the following, we define the strong confinement regime as the case, where the phonon frequency $\nu$ is comparable or larger than the spontaneous cavity decay rate $\kappa$. In this subsection we therefore assume that \footnote{Notice that we do not restrict ourselves here to the case, where $\nu \gg \kappa$, as it is usually done \cite{Tony2}. This means, we define the strong confinement regime here in a more generous way.}
\begin{eqnarray} \label{cond}
\kappa < \nu , \, \delta_{\rm eff} ~~ {\rm and} ~~ \eta \ll 1\, .
\end{eqnarray}
In this parameter regime, the time scale separation which we assumed in the previous subsection no longer applies. This means, a proper analysis of the cooling process should take all cooling equations into account. However, as we shall see below, the cooling process takes place on a time scale which is much longer than the time scale given by the inverse cavity decay rate $1/ \kappa$. This means, $n_2$ evolves only on a much longer time scale than all the other above defined expectation values. It is therefore possible to simplify the 25 cooling equations introduced in this paper again via an adiabatic elimination. The details of this calculation can be found in App.~\ref{appC}, where we calculate $n_1$, $k_{11}$, and $k_{12}$ up to zeroth and first order in $\eta$, respectively.

\begin{figure}[t]
\begin{minipage}{\columnwidth}
\includegraphics[scale=0.9]{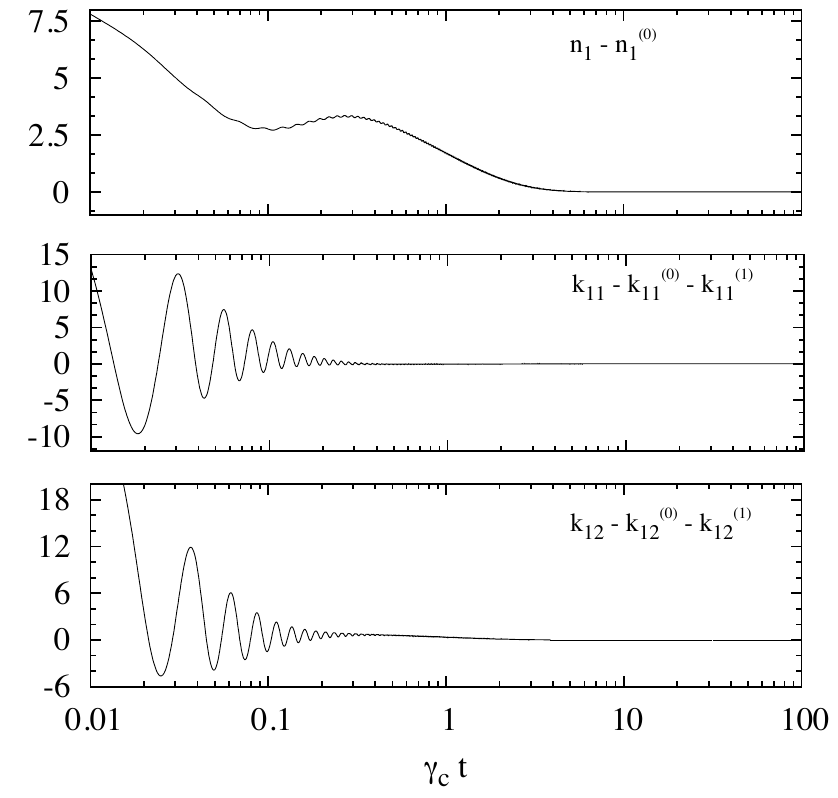}
\caption{Difference between the analytical and the numerical solutions for $n_1$, $k_{11}$, and $k_{12}$ as in Fig.~\ref{pic} but for $\eta=0.1$, $\nu = 10 \, \kappa$, $\delta_{\rm eff}  = \nu$, and $g_{\rm eff} = \kappa$ which are typical experimental parameters for a {\em strongly} coupled cavity in the {\em strong} confinement regime.} \label{pic4}
\end{minipage}
\end{figure}

Figs.~\ref{pic3} and \ref{pic4} compare the obtained analytical results with the corresponding numerical solutions of the closed set 25 cooling equations. In case of a relatively weakly coupled optical cavity (with $g_{\rm eff} \ll \kappa$) we find again relatively good agreement between both solutions. Although we now eliminate more variables from the system dynamics, we find again that the results of the adiabatic elimination apply to a very good agreement throughout the whole cooling process. Less agreement is found in the case of a strongly coupled optical cavity with $g_{\rm eff} = \kappa$. In this case, the expressions found for $n_1$, $k_{11}$, and $k_{12}$ apply only towards the end of the cooling process. Unfortunately, it is not possible to obtain more accurate for this parameter regimes and the case where $g_{\rm eff} \gg \kappa$. This would require to calculate $n_1$, $k_{11}$, and $k_{12}$ up to terms in the order of $\eta^2$ correctly which is beyond the possible scope of this paper.  

Substituting Eqs.~(\ref{zeroth}) and (\ref{k111}) into Eq.~(\ref{4888}), we now obtain only a single effective cooling equation,
\begin{eqnarray} \label{strong2}
\dot n_2 &=& - \gamma_{\rm c} \, n_2 + c 
\end{eqnarray}
with the constants $\gamma_{\rm c}$ and $c$ given by
\begin{eqnarray} \label{eff88strong}
\gamma_{\rm c} &=& {64 \eta^2 \kappa \nu \delta_{\rm eff} g_{\rm eff}^2 \over \left[\kappa^2 + 4 (\delta_{\rm eff} + \nu)^2 \right] \left[\kappa^2 + 4 (\delta_{\rm eff} - \nu)^2 \right]} \, , \nonumber \\
c &=&{4 \eta^2 \kappa g_{\rm eff}^2 \over \kappa^2 + 4(\delta_{\rm eff} + \nu)^2} 
\end{eqnarray}
up to second order in $\eta$. As we shall see below, $\gamma_{\rm c}$ is the effective cooling rate for the cavity mediated cooling process illustrated in Fig.~\ref{setup}.

In zeroth order in $\eta$, there is no difference between $n_2$ and the mean phonon number $m$ (cf.~Eq.~(\ref{mfinaldot})). Eq.~(\ref{strong2}) is hence identical to the effective cooling equation (\ref{strong24}). A comparison between both equations shows that the rates $A_\pm$ equal
\begin{eqnarray} \label{eff88strongx}
A_\pm &=& {4 \kappa g_{\rm eff}^2 \over \kappa^2 + 4(\delta_{\rm eff} \pm \nu)^2} \, .
\end{eqnarray}
These expressions for the rates $A_\pm$ are consistent with the analogous expressions obtained in Ref.~\cite{Cirac4,morigi,morigi2}. As we shall see below in Section \ref{progress}, Eqs.~(\ref{strong2}) and (\ref{eff88strong}) 
--- and therefore also Eq.~(\ref{eff88strongx}) --- apply in the weak as well as in the strong confinement regime.

\section{Stability analysis} \label{stability}

\noindent \begin{figure*}[t]
\begin{minipage}{2\columnwidth}
\begin{center}
\resizebox{\columnwidth}{!}{\rotatebox{0}{\includegraphics{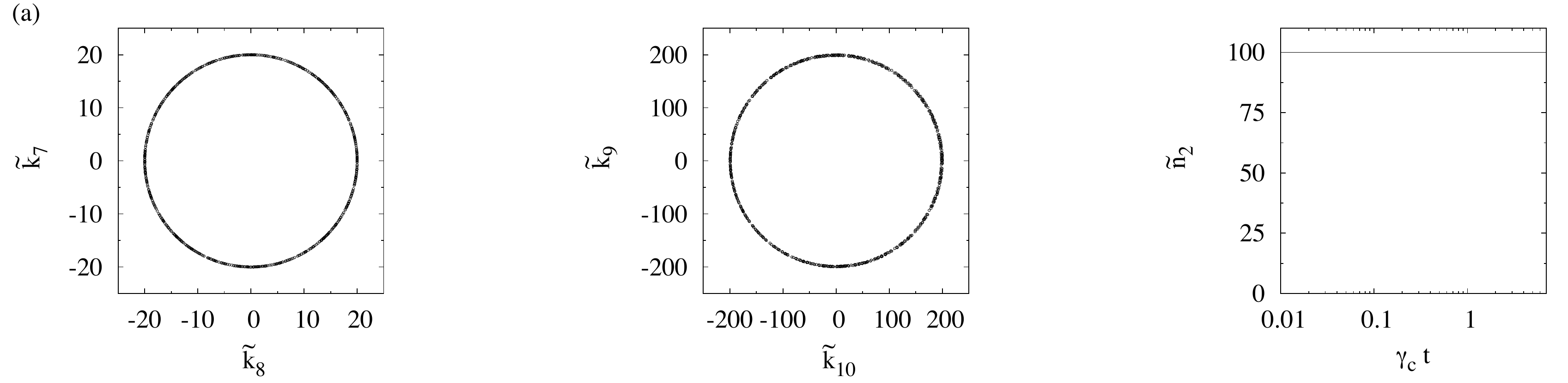}}} 
\resizebox{\columnwidth}{!}{\rotatebox{0}{\includegraphics{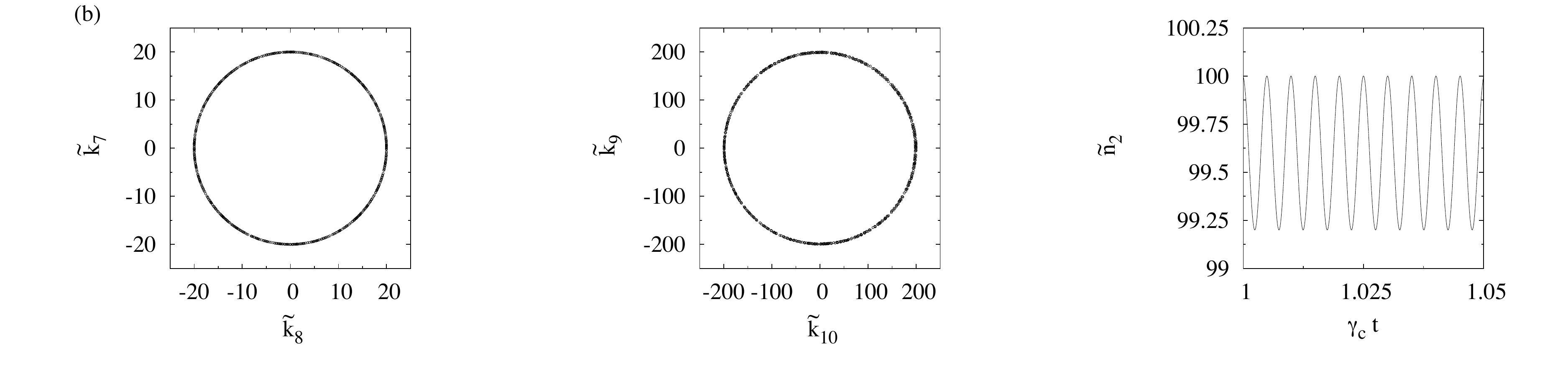}}}
\resizebox{\columnwidth}{!}{\rotatebox{0}{\includegraphics{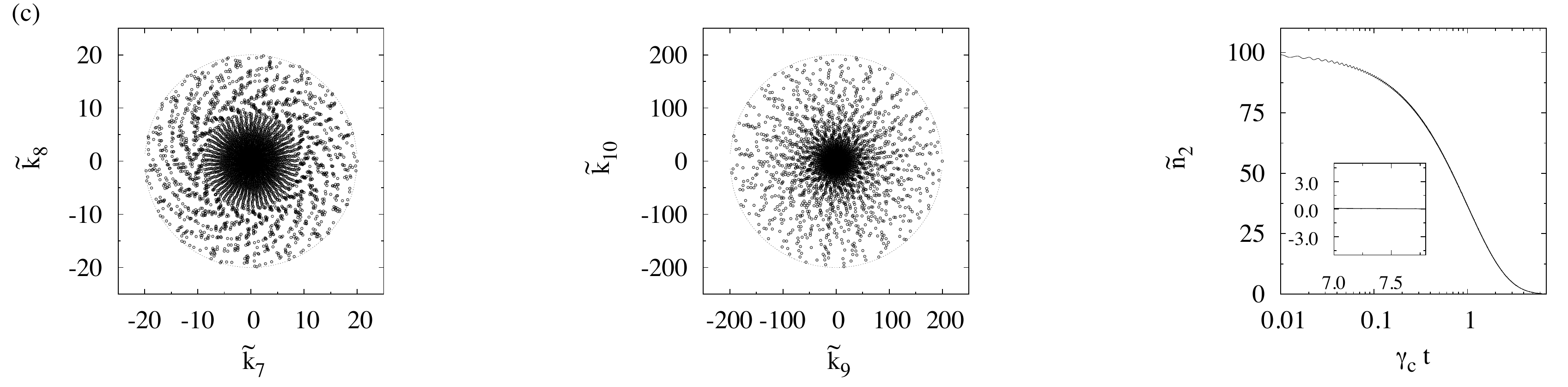}}}
\end{center}
\caption{Diagrams illustrating the time evolution of the expectation values $\tilde k_7$ to $\tilde k_{10}$, and $\tilde n_2$ for $\eta = 0.1$, $\nu = 0.1 \, \kappa$, $\delta_{\rm eff} = 0.5 \, \kappa$, $g_{\rm eff} = 0.1 \, \kappa$ and $\tilde n_2 (0) = 100$. All figures are the result of a numerical solution of the effective cooling equations in Eq.~(\ref{eff}). In (a), only terms in zeroth order in $\eta$ are taken into account. In (b), terms in zeroth and in first order in $\eta$ are taken into account. In (c), all matrix elements of $M$ in Eq.~(\ref{eff}) are taken into account.} \label{eta_zero}
\end{minipage}
\end{figure*}

In the strong confinement regime (cf.~Eq.~(\ref{strong2})), the cooling process can be described by a single effective cooling equation. Since the cooling rate $\gamma_{\rm c}$ is always positive, the trapped particle always reaches its stationary state. However, it is not clear whether or not the same applies in the weak confinement regime, where the cooling process is described by five linear differential equations (cf.~Eq.~(\ref{eff})). Proceeding as in Ref.~\cite{Tony2}, we now have a closer look at the dynamics induced by these equations. To do so, we introduce the shifted $y$ operator expectation values 
\begin{eqnarray} \label{effxx}
\big( \tilde n_2 , \tilde k_7 , \tilde k_8 , \tilde k_9 , \tilde k_{10} \big)^{\rm T} 
&\equiv & \left( n_2 , k_7 , k_8 , k_9 , k_{10} \right)^{\rm T} \nonumber \\
&& + M^{-1} \left( \beta_1 , \beta_2 , \beta_3 , \beta_4 , \beta_5 \right)^{\rm T} . ~~~~~~
\end{eqnarray}
Notice that the tilde and the non-tilde variables differ only by constants, namely by the stationary state solutions of the non-tilde expectation values. Substituting Eq.~(\ref{effxx}) into the effective cooling equations in Eq.~(\ref{eff}), they hence simplify to
\begin{eqnarray} \label{effyy}
\big( \dot{\tilde{n}}_2 , \dot{\tilde k}_7 , \dot{\tilde k}_8 , \dot{\tilde k}_9 , \dot{\tilde k}_{10} \big)^{\rm T} 
&=& M \left( \tilde n_2 , \tilde k_7 , \tilde k_8 , \tilde k_9 , \tilde k_{10} \right)^{\rm T} . ~~~~
\end{eqnarray}
The stationary state solution of this differential equation is the trivial one with all tilde variables equal to zero. In the following we show that the real parts of all eigenvalues of $M$ are negative, which is a necessary condition for the system to reach this state. 

\subsection{Time evolution for $\eta = 0$}

First, we calculate the eigenvalues of $M$ in Eq.~(\ref{eff3}) for $\eta = 0$ and find that these are simply given by 
\begin{eqnarray} \label{eff55}
\lambda_1 = 0 \, , ~~ \lambda_{2,3} = \mp {\rm i} \nu \, , ~~ \lambda_{4,5} = \mp 2 {\rm i} \nu \, .  
\end{eqnarray}
Taking this into account and solving Eq.~(\ref{effyy}) analytically, we find that
\begin{eqnarray} \label{eff552}
\tilde n_2 (t) &=& \tilde n_2(0) \, , \nonumber \\  
\left( \begin{array}{c} \tilde k_7 (t) \\ \tilde k_8 (t) \end{array} \right) &=& \left( \begin{array}{rr} \cos \nu t & - \sin \nu t \\ \sin \nu t & \cos \nu t \end{array} \right) \left( \begin{array}{c} \tilde k_7 (0) \\ \tilde k_8 (0) \end{array} \right) \, , \nonumber \\
\left( \begin{array}{c} \tilde k_9 (t) \\ \tilde k_{10} (t) \end{array} \right) &=& \left( \begin{array}{rr} \cos 2 \nu t & - \sin 2 \nu t \\ \sin 2 \nu t & \cos 2 \nu t \end{array} \right) \left( \begin{array}{c} \tilde k_9 (0) \\ \tilde k_{10} (0) \end{array} \right) . ~~~
\end{eqnarray}
These equations are illustrated in Fig.~\ref{eta_zero}(a) which shows phase diagrams for the time evolution of the coherences $\tilde k_7$ to $\tilde k_{10}$. The fact that all points lie on a circle illustrates that an initially coherent state of the $y$ particles remains essentially coherent throughout the cooling process.  The numerical solution for the time evolution of $n_2$ shows that, for $\eta = 0$, the mean phonon number $m$ does not change in time, as one would expect. There cannot be any cooling without an interaction between the electronic and the motional states of the trapped particle. 

\subsection{First order corrections}

Calculating the eigenvalues of the matrix $M$ in Eq.~(\ref{eff3}) up to first order in $\eta$, we obtain again Eq.~(\ref{eff55}). All of them have zero real parts. But there are first order corrections to the eigenvectors of $M$. As a result, $\tilde n_2$ is no longer constant in time. This is illustrated in Fig.~\ref{eta_zero}(b) which shows a numerical solution of Eq.~(\ref{effyy}) with all first order corrections in $\eta$ taken into account. However, since the eigenvalues of $M$ have no real parts, $\tilde n_2$ and therefore also the mean phonon number $m$, do not reach their stationary state solutions. Instead, $\tilde n_2$ remains close to its initial value. No cooling occurs.

\subsection{Second order corrections}

Taking all terms in Eq.~(\ref{eff3}) into account, one can show that the eigenvalues of $M$ are without any approximations given by
\begin{eqnarray} 
\lambda_1 &=& \alpha_{11}^{(2)} \, , \notag \\
\lambda_{2,3} &=& {1\over 2} \, \alpha_{11}^{(2)} \mp {{\rm i} \over 2} \sqrt { 4 \nu^2 - \alpha_{11}^{(2)\, 2}} \, , \notag \\
\lambda_{4,5} &=& \alpha_{11}^{(2)} \mp {\rm i} \sqrt{ 4 \nu^2 - \alpha_{14}^{(2)} \alpha_{41}^{(2)}} \, .
\end{eqnarray}
For positive effective laser detunings, the matrix element $\alpha_{11}^{(2)}$ (cf.~Eq.~(\ref{eff6})) is always negative. This means, all eigenvalues of $M$ have negative real parts, when $\delta_{\rm eff} > 0$. In this case, all tilde variables are damped away on the time scale given by $1/\alpha_{11}^{(2)}$ and tend eventually to zero. This is illustrated in Fig.~\ref{eta_zero}(c). Now we observe an exponential damping of $\tilde n_2$ which implies cooling of the mean number of phonons $m$. Analogously, one would find heating when solving the above equations for negative effective laser detunings, ie.~$\delta_{\rm eff} < 0$.

Moreover, for $\delta_{\rm eff} > 0$, we find that the $y$ coherences $\tilde k_7$ to $\tilde k_{10}$ oscillate with a slowly decreasing amplitude around zero. Analogously one can show that the $y$ coherences $k_7$ to $k_{10}$ oscillate with a slowly decreasing amplitude around their time averages. This means, the cooling process remains stable and the trapped particle can be expected to reach its stationary state eventually. This observation is taken into account in the following section, where we analyse the cooling process in more detail by replacing the coherences $k_7$ to $k_{10}$ by their time averages. 

\section{Phonon numbers and cooling rates} \label{progress}

In this section, we point out that the effective cooling equation for $n_2$ in Eq.~(\ref{strong2}) applies to a very good approximation not only in the strong confinement regime but also in the weak confinement regime. Since $n_2$ and the mean phonon number $m$ are identical in zeroth order in $\eta$, solving this equation implies that $m$ is to a very good approximation given by
\begin{eqnarray} \label{mfinaldot4}
m (t) &= & \left[ m(0) - m_{\rm ss}  \right] \, {\rm e}^{ - \gamma_{\rm c} t} + m_{\rm ss} \, ,
\end{eqnarray}
with $\gamma_{\rm c}$ as in Eq.~(\ref{eff88strong}) and with $m_{\rm ss}$,  
\begin{eqnarray} \label{mss0}
m_{\rm ss} &=& {c \over \gamma_{\rm c}} \, ,
\end{eqnarray}
being the stationary state phonon number for the cooling process illustrated in Fig.~\ref{setup} in zeroth order in $\eta$. 

\subsection{Effective time evolution} \label{average}

The previous section shows that, in the weak confinement regime, the initial $y$ operator coherences $k_7$ to $k_{10}$ oscillate relatively rapidly in time. However, since they oscillate with a decreasing amplitude, we can safely approximate them by their time averages. The easiest way of calculating these time averages is to recognise that their time derivatives are equal to zero. This means, the time averages of $k_7$ to $k_{10}$ are the solutions of 
\begin{eqnarray} \label{so}
\dot k_i &=& 0 ~~ {\rm for} ~~ i=7,...,10 \, .
\end{eqnarray}
Exactly the same condition has been imposed in Section \ref{effeq2} and App.~\ref{appC}, when analysing the time evolution of $n_2$ in the strong confinement regime via an adiabatic elimination of $k_7$ to $k_{10}$. This means, the calculations in Section \ref{effeq2}, and therefore also Eq.~(\ref{strong2}), apply also in the weak confinement regime to a very good approximation. 

\subsection{Stationary state phonon number}

\begin{figure}[t]
\begin{minipage}{\columnwidth}
\includegraphics[scale=0.9]{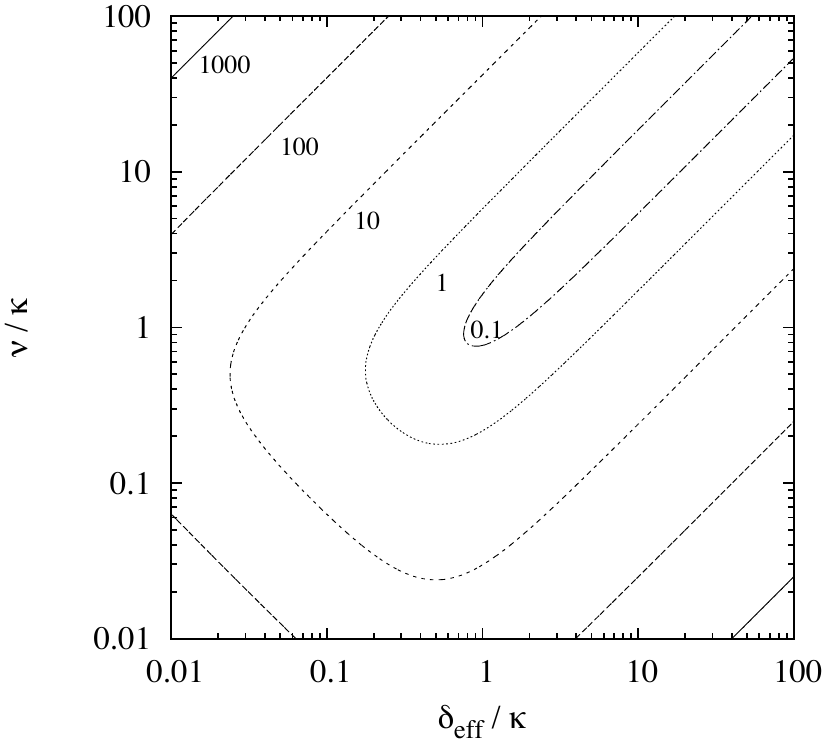}
\caption{Logarithmic contour plot of the stationary state phonon number $m_{\rm ss}$ in Eq.~(\ref{mss}) for a strongly coupled optical cavity as a function of the phonon frequency $\nu$ and the effective laser detuning $\delta_{\rm eff}$.} \label{c2}
\end{minipage}
\end{figure}

Substituting Eq.~(\ref{eff88strong}) into Eq.~(\ref{mss0}), we find that the stationary state phonon number $m_{\rm ss}$ is in  zeroth order in $\eta$ given by
\begin{eqnarray} \label{mss}
m_{\rm ss} &=& {\kappa^2 + 4 (\delta_{\rm eff} - \nu)^2 \over 16 \nu \delta_{\rm eff}} \, . 
\end{eqnarray}
That this term is exactly the same as the stationary state phonon number obtained by other authors (cf.~eg.~Ref.~\cite{Tony}), shows that our calculations are consistent with previous calculations. For example, in the weak confinement regime (cf.~Eq.~(\ref{oups})), the stationary state phonon number $m_{\rm ss}$ assumes its minimum, when  
\begin{eqnarray} \label{old2}
\delta_{\rm eff} &=& {1 \over 2} \, \kappa \, .
\end{eqnarray}
As already pointed out in Ref.~\cite{Tony}, this detuning corresponds to the stationary state phonon number
\begin{eqnarray} \label{mssweak}
m_{\rm ss} &=& {\kappa \over 4 \nu}  
\end{eqnarray}
which is in general much larger than one. In the strong confinement regime (cf.~Eq.~(\ref{cond})), the stationary state phonon number $m_{\rm ss}$ assumes its minimum, when  
\begin{eqnarray} \label{old}
\delta_{\rm eff} &=& {1 \over 2} \, \sqrt{\kappa^2 + 4 \nu^2} \, .
\end{eqnarray}
For spontaneous decay rates $\kappa$ much smaller than $\nu$, this equation simplifies to $\delta_{\rm eff} = \nu$ (sideband cooling). Substituting this result into Eq.~(\ref{mss}) and assuming $\kappa \ll \nu$, we see that the minimum stationary state phonon number $m_{\rm ss}$ equals 
\begin{eqnarray} \label{mssstrong}
m_{\rm ss} &=& {\kappa^2 \over 16 \nu^2}  
\end{eqnarray}
in this case which is indeed much smaller than one. These results are confirmed by Fig.~\ref{c2}, which shows $m_{\rm ss}$ as a function of $\nu/\kappa$ and $\delta_{\rm eff}/\kappa$.

\subsection{Effective cooling rate} 

\begin{figure}[t]
\begin{minipage}{\columnwidth}
\includegraphics[scale=0.9]{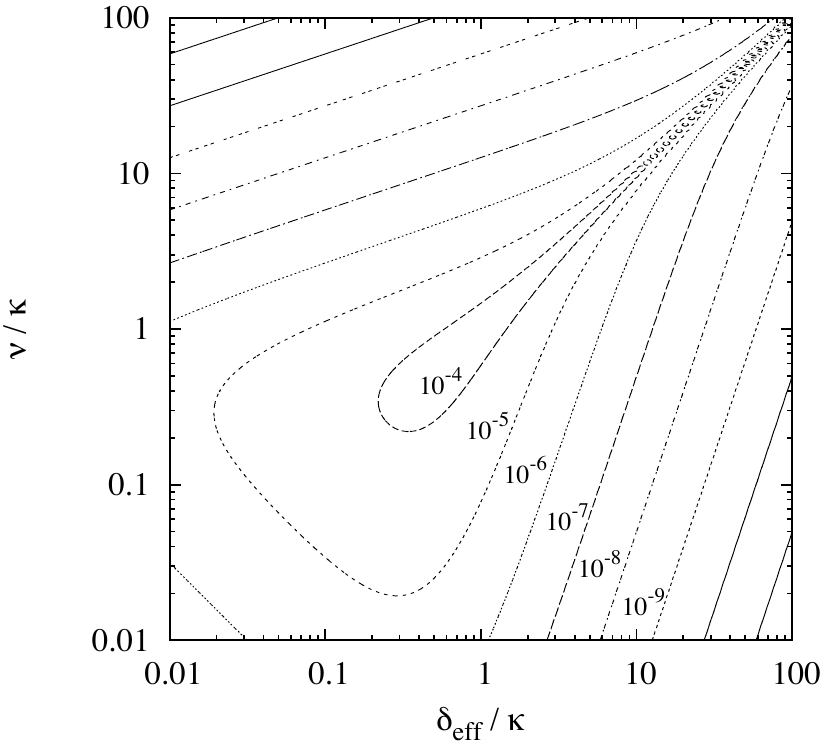}
\caption{Logarithmic contour plot of the effective cooling rate $\gamma_{\rm c}$ in Eq.~(\ref{eff88strong}) in units of $2 g_{\rm eff}^2/\kappa$ as a function of the relative phonon frequency $\nu/\kappa$ and the relative effective detuning $\delta_{\rm eff}/\kappa$ for $\eta = 0.01$.} \label{c3}
\end{minipage}
\end{figure}

Let us now have a closer look at typical values of the effective cooling rate $\gamma_{\rm c}$. Fig.~\ref{c3} shows $\gamma_{\rm c}$ in units of $2 g_{\rm eff}^2/\kappa$ as a function of $\nu/\kappa$ and $\delta_{\rm eff}/\kappa$. Since $\gamma_{\rm c}$ always scales as $g_{\rm eff}^2$, the cooling rate $\gamma_{\rm c}$ might be very small for realistic experimental parameters. In this case, it might seem as if the system reaches its stationary state, even when $m_{\rm ss}$ is very small.

\subsection{Numerical results} \label{num}

\begin{figure*}[t]
\begin{minipage}{2\columnwidth}
\begin{center}
\resizebox{\columnwidth}{!}{\rotatebox{0}{\includegraphics{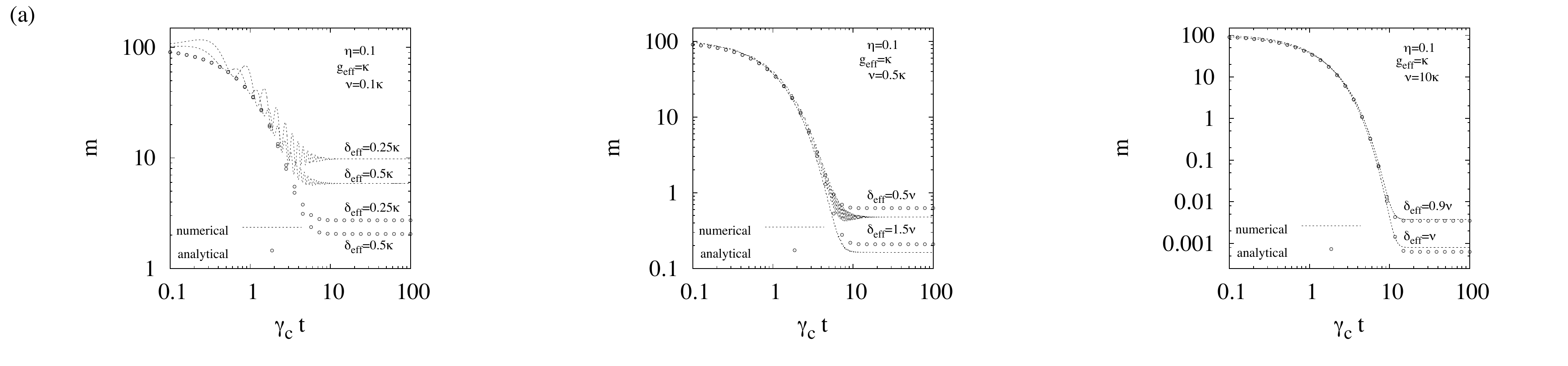}}} 
\resizebox{\columnwidth}{!}{\rotatebox{0}{\includegraphics{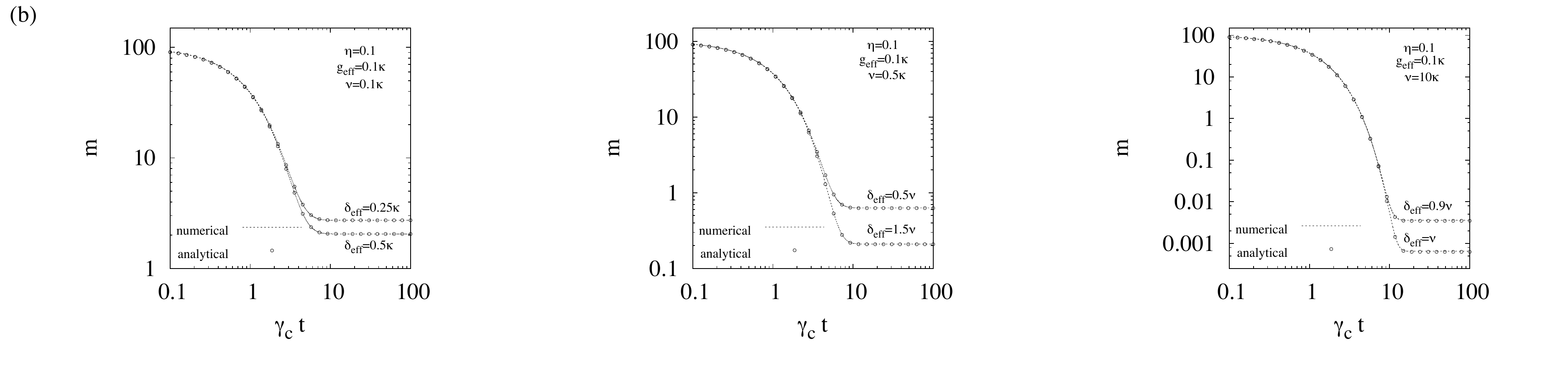}}}
\end{center}
\caption{Logarithmic plots of the time evolution of the mean phonon number $m$ during the cooling process for typical experimental parameters in the {\em strong} confinement regime. The dashed lines are the result of a numerical integration of the 25 cooling equations which can be found in Section \ref{EV} and in App.~\ref{appB}. The circles illustrate the analytical solution given in Eq.~(\ref{mfinaldot4}).} \label{time}
\end{minipage}
\end{figure*}

We conclude this section with a numerical solution of the full set of 25 cooling equations which we can be found in this paper in Section \ref{standard} and App.~\ref{appB}. Fig.~\ref{time}(a) illustrates the cooling process for a relatively strongly coupled cavity with $g_{\rm eff} = \kappa$. Fig.~\ref{time}(b) illustrates the cooling process for a weakly coupled cavity with $g_{\rm eff} \ll \kappa$. We then compare these solutions with our analytical solution for the time evolution of the mean phonon number $m$ which takes the effective cooling rate $\gamma_{\rm c}$ in Eq.~(\ref{eff88strong}) and the stationary state phonon number $m_{\rm ss}$ in Eq.~(\ref{mss}) into account. 

\noindent 

A closer look at Fig.~\ref{time} confirms that there is very good agreement between analytical and numerical results, in the case of a weakly coupled cavity. In the case of a strongly coupled cavity, we only observe reasonable agreement in the strong confinement regime when $\nu > \kappa$. However, when modelling the cooling process for a weakly confined particle inside a strongly coupled cavity, we find that the analytical expression for the stationary state phonon number $m_{\rm ss}$ in Eq.~(\ref{mss}) is substantially lower than the corresponding numerical solution. This difference tells us that higher order terms in $\eta$ should be taken into account when calculating $n_1$, $k_{11}$, and $k_{12}$ via an adiabatic elimination, as pointed out already in Section \ref{effeq2}. A much larger set of more accurate rate equations should be taken into account.

\section{Conclusions} \label{conc}

This paper revisits a standard scenario for cavity mediated laser cooling \cite{Cirac2,Cirac4,cool,morigi,morigi2,Tony}. As illustrated in Fig.~\ref{setup}, we consider a particle, an atom, ion, or molecule, with ground state $|0 \rangle$ and excited state $|1 \rangle$ with an external trap inside an optical cavity. Moreover, we assume that the motion of the particle orthogonal to the cavity axis, ie.~in the direction of the cooling laser, is either strongly or weakly confined and consider it quantised. The cooling laser establishes a direct coupling between the phonons and the electronic states of the trapped particle, thereby resulting in the continuous conversion of phonons into cavity photons. When these leak into the environment, vibrational energy is permanently lost from the system which implies cooling. 

As in Refs.~\cite{Cirac2,Cirac4,cool,morigi,morigi2,Tony}, we describe the time evolution of the experimental setup in Fig.~\ref{setup} with the help of a quantum optical master equation. Assuming that the excited state $|1 \rangle$ of the trapped particle is strongly detuned (c.f.~Eq.~(\ref{condi0})), the system Hamiltonian can be simplified via an adiabatic elimination of the electronic states of the trapped particle. We then use the resulting effective master equation to obtain a closed set of 25 rate equations, ie.~linear differential equations, which describe the time evolution of expectation values. Most of these expectation values are coherences.

Since the effective cooling rate $\gamma_{\rm c}$ (cf.~Eq.~(\ref{eff88strong})) scales as $\eta^2$, a proper analysis of the cooling process needs to take terms of the order $\eta^2$ in the system dynamics into account. Instead of expanding the Hamiltonian $H_{\rm I}$ in Eq.~(\ref{2.8}) in $\eta$, we solve the cooling equations for small Lamb-Dicke parameters $\eta$ perturbatively. The reason that our calculations are nevertheless relatively straightforward is that we replace the phonon and the cavity photon annihilation operators $b$ and $c$ in the interaction Hamiltonian $H_{\rm I}$ by two new bosonic operators $x$ and $y$  (c.f.~Eqs.~(\ref{25}) and (\ref{defy})) which describe the cavity-phonon system in a more natural way and commute with each other (c.f.~Eq.~(\ref{33})). The operator $x$ annihilates cavity photons while giving a kick to the trapped particle. The operator $y$ annihilates phonons but not without affecting the field inside the optical cavity. 

Our results confirm that there are many similarities between ordinary and cavity mediated laser cooling \cite{Tony}. However, for a weakly confined particle inside a strongly coupled cavity, a comparison between analytical and numerical results suggests that more detailed calculations are needed to model the cooling process accurately. Our analytical calculations are designed such that they calculate $m_{\rm ss}$ in zeroth order in the Lamb-Dicke parameter $\eta$ (cf.~Eq.~(\ref{mss})). This means, we neglect higher order terms in $\eta$ in the rate equations, whenever possible. Our numerical calculations however take all available terms in the above mentioned 25 rate equations into account. The difference between analytical and numerical results means that terms of higher order in $\eta$ are not negligible, although this might seem to be the case. Unfortunately, calculating $m_{\rm ss}$ systematically up to first order in $\eta$, either analytically or numerically, would require to take considerably more than only 25 cooling equations into account.

The above observation is nevertheless interesting, since the cooling of a weakly confined particle inside a strongly coupled cavity is of practical interest for the cooling of molecules. Realising a very strong coupling between a trapped particle and the field inside an optical cavity is in principle feasible \cite{Kim3,Kim4}. Over the last years, experiments have been performed with a continuously increasing ratio between the cavity coupling constant $g$ and the spontaneous cavity decay rate $\kappa$. Even larger ratios $g/\kappa$ are expected to occur when trapping large molecules, whose electric dipole moment ${\bf D}_{01}$ can be much larger than that of an atom, inside an optical cavity. Such molecules can experience a relatively large effective cavity coupling constant $g_{\rm eff}$. \\[0.5cm]

{\em Acknowledgement.} The authors would like to thank Philippe Grangier and Giuseppe Vitiello for stimulating discussions and many helpful comments. This work was supported by the UK Research Council EPSRC.

\appendix
\section{Adiabatic elimination of the electronic states} \label{adel}

In the following, we write the state vector of the atom-cavity-phonon system as
\begin{eqnarray}
|\psi\rangle=  \sum_{j=0}^1 \sum_{m=0}^\infty \sum_{n=0}^\infty c_{jmn} \, |jmn \rangle \, ,
\end{eqnarray} 
where $|j \rangle$ and $|m \rangle$ denote the electronic and the vibrational energy eigenstates of the particle and where $|n \rangle$ is a cavity photon number state. According to the Schr\"odinger equation, the time evolution of the coefficient $c_{j'm'n'}$ is given by
\begin{eqnarray}\label{3.3}
\dot{c}_{j',m',n'} &=& - {{\rm i} \over \hbar} \sum_{j=0}^1 \sum_{m=0}^\infty \sum_{n=0}^\infty c_{jmn} \, \langle j'm'n' | H_{\rm I} |jmn \rangle \, . ~ \nonumber \\
\end{eqnarray}
Given condition (\ref{condi0}), the coefficients $c_{j'm'n'}$ with $j' = 1$ evolve on a much faster time scale than the coefficients with $j' = 0$. Setting the time derivatives of these coefficients equal to zero, we find that
\begin{eqnarray} \label{c1-c0} 
c_{1m' n'} &=& - \frac{1}{2 \Delta} \sum_{m=0}^\infty \sum_{n=0}^\infty c_{0mn} \notag \\ 
&&  \times \langle m'n' | \, \left( \Omega \, D^{\dag} ({\rm i} \eta) + 2 g \, c \right) \, | mn  \rangle \, ,  
\end{eqnarray}
if the particle is initially in its ground state. This equation holds up to first order in $1/\Delta$. Substituting this result into Eq.~(\ref{3.3}) and neglecting an overall level shift, we obtain the effective interaction Hamiltonian in Eq.~(\ref{3.7}).

\section{Relevant expectation values} \label{appA}

The calculations in Apps.~\ref{appB} and \ref{appC} require in addition to the expectation values defined in Section \ref{EV} the $x$ operator expectation values
\begin{eqnarray} \label{coherencesappA}
&& \hspace*{-0.4cm} k_1 \equiv \langle x + x^\dagger \rangle \, , ~~
k_2 \equiv {\rm i} \, \langle x - x^\dagger \rangle \, , \notag \\
&& \hspace*{-0.4cm} k_3 \equiv  \langle x^2 + x^{\dagger 2} \rangle \, , ~~
k_4 \equiv  {\rm i}  \,  \langle x^2 - x^{\dagger 2} \rangle \, , \notag \\
&& \hspace*{-0.4cm} k_5 \equiv \langle x^\dagger (x + x^\dagger ) x \rangle \, , ~~
k_6 \equiv {\rm i}  \, \langle x^\dagger (x - x^\dagger ) x \rangle \, .
\end{eqnarray}
Moreover we employ in the following the mixed operator expectation values $k_{15}$ to $k_{22}$ which are defined as
\begin{eqnarray} \label{coherences3}
k_{13} &\equiv & \langle (x + x^\dagger ) y^\dagger y \rangle \, , \notag \\
k_{14} &\equiv & {\rm i}  \, \langle (x - x^\dagger ) y^\dagger y \rangle \, , \notag \\
k_{15} &\equiv & \langle (x - x^\dagger ) (y - y^\dagger ) \rangle \, , \notag \\
k_{16} &\equiv & {\rm i} \, \langle(x + x^\dagger ) (y - y^\dagger ) \rangle \, , \notag \\
k_{17} &\equiv & \langle (x + x^\dagger ) (y + y^\dagger ) \rangle \, , \notag \\
k_{18} &\equiv & {\rm i} \, \langle(x - x^\dagger ) (y + y^\dagger ) \rangle \, , \notag \\
k_{19} &\equiv & \langle (x - x^\dagger) (y^2 - y^{\dagger 2}) \rangle \, , \nonumber \\ 
k_{20} &\equiv & {\rm i} \, \langle (x + x^\dagger) (y^2 - y^{\dagger 2} ) \rangle \, , \nonumber \\ 
k_{21} &\equiv & \langle (x + x^\dagger) (y^2 + y^{\dagger 2} ) \rangle \, , \nonumber \\ 
k_{22} &\equiv & {\rm i} \, \langle (x - x^\dagger) (y^2 + y^{\dagger 2} ) \rangle \, .   
\end{eqnarray}
Their time derivatives of these and other expectation values can be found in App.~\ref{appB}, where they are use to obtain a reduced set of effective cooling equations.

\section{$n_1$, $k_{11}$, and $k_{12}$ in the weak confinement regime} \label{appB}

In this appendix, we derive approximate solutions for the expectation values $k_{11}$, $k_{12}$, and $n_1$ for the weak confinement regime (cf.~Eq.~(\ref{oups})). This is done via an adiabatic elimination of the $x$ and the mixed operator expectation values which all evolve on the relatively fast time scale given by the spontaneous cavity decay rate $\kappa$.   
To indicate the scaling of variables, we adopt the notation  
\begin{eqnarray}
x= x^{(0)} + x^{(1)} + x^{(2)} + ... 
\end{eqnarray}
The superscripts indicate the scaling of the respective terms with respect to $\eta$. As we shall see below, the expectation values $k_{11}$, $k_{12}$, and $n_1$ need to be calculated up to first order in $\eta$. Let us first have a look at $k_{11}^{(0)} $, $k_{12}^{(0)} $, and $n_1^{(0)} $.

Using Eq.~(\ref{dotA0}) and setting $\eta = 0$, we find that $n_1$, $n_3$, and $k_1$ to $k_6$ evolve in zeroth order in $\eta$ according to
\begin{eqnarray} \label{48882}
\dot n_1 &=& g_{\rm eff} \, k_2 - \kappa \, n_1  \, , \nonumber \\
\dot n_3 &=& g_{\rm eff} \, \left( k_2 + 2 k_6 \right) + \kappa \, \left( n_1 - 2 n_3 \right) \, , \nonumber \\
\dot k_1 &=& - \delta_{\rm eff} \, k_2 - {1 \over 2} \kappa \, k_1 \, , \nonumber \\ 
\dot k_2 &=& 2 g_{\rm eff} + \delta_{\rm eff} \, k_1 -  {1 \over 2} \kappa \, k_2 \, , \nonumber \\
\dot k_3 &=& - 2 g_{\rm eff} \, k_2 - 2 \delta_{\rm eff} \, k_4  - \kappa \, k_3 \, , \nonumber \\
\dot k_4 &=& 2 g_{\rm eff} \, k_1 + 2 \delta_{\rm eff} \, k_3 - \kappa \, k_4  \, , \nonumber \\
\dot k_5 &=& g_{\rm eff} \, k_4 - \delta_{\rm eff} \, k_6 - {3 \over 2} \kappa \, k_5 \, , \nonumber \\
\dot k_6 &=& g_{\rm eff} \, \left( 4 n_1 - k_3 \right) + \delta_{\rm eff} \, k_5 - {3 \over 2} \kappa  \, k_6 \, . ~~~~
\end{eqnarray}
These equations form a closed set of differential equations. Eliminating the above $x$-operator expectation values adiabatically from the system dynamics, we find for example that $n_1$ is in zeroth order in $\eta$ given by
\begin{eqnarray} \label{zeroth}
n_1^{(0)} &=& {4 g_{\rm eff}^2 \over \kappa^2  + 4 \delta_{\rm eff}^2} \, . ~~
\end{eqnarray}
In addition we obtain expressions for $k_1^{(0)}$, $k_2^{(0)}$, $k_5^{(0)}$, $k_6^{(0)}$, and $n_3^{(0)}$. These are used later on in this appendix to calculate $k_{11}^{(1)}$ and $k_{12}^{(1)}$.

Setting $\eta = 0$ and using again Eq.~(\ref{dotA0}), we moreover find that the time evolution of the mixed operator coherences $k_{11}$ and $k_{12}$ and $k_{15}$ to $k_{18}$ is in zeroth order in $\eta$ is given by
\begin{eqnarray} \label{7772}
\dot k_{11} &=& g_{\rm eff} \, k_{18} - \nu \, k_{12} - \kappa \, k_{11} \, , \nonumber \\ 
\dot k_{12} &=& - g_{\rm eff} \, k_{15} + \nu \, k_{11} - \kappa \, k_{12} \, , \nonumber \\
\dot k_{15} &=& - 2 g_{\rm eff} \, k_8 - \delta_{\rm eff} \, k_{16} - \nu \, k_{18} - {1 \over 2} \kappa \, k_{15} \, , \nonumber \\
\dot k_{16} &=& \delta_{\rm eff} \, k_{15} + \nu \, k_{17} - {1 \over 2} \kappa \, k_{16} \, , \nonumber \\
\dot k_{17} &=& - \delta_{\rm eff} \, k_{18} - \nu \, k_{16} - {1 \over 2} \kappa \, k_{17} \, , \nonumber \\
\dot k_{18} &=& 2 g_{\rm eff} \, k_7 + \delta_{\rm eff} \, k_{17} + \nu \, k_{15} - {1 \over 2} \kappa \, k_{18} \, .
\end{eqnarray}
These six equations too form a closed set of cooling equations which describe a time evolution on the time scale of the spontaneous cavity decay rate $\kappa$. Taking this into account, eliminating $k_{11}$ and $k_{12}$ and $k_{15}$ to $k_{18}$ adiabatically, and neglecting terms proportional to $\nu^2$ which are much smaller than the remaining terms, we find that
\begin{eqnarray} \label{k11120}
k_{11}^{(0)} &=& {4 g_{\rm eff}^2 \over \kappa^2 + 4 \delta_{\rm eff}^2} \, k_7 - {4 \nu g_{\rm eff}^2 ( 3 \kappa^2 - 4 \delta_{\rm eff}^2 ) \over \kappa (\kappa^2 + 4 \delta_{\rm eff}^2)^2} \, k_8 \, , \notag\\
k_{12}^{(0)} &=& {4 \nu g_{\rm eff}^2 ( 3 \kappa^2 - 4 \delta_{\rm eff}^2 ) \over \kappa (\kappa^2 + 4 \delta_{\rm eff}^2)^2} \, k_7 +  {4 g_{\rm eff}^2 \over \kappa^2 + 4 \delta_{\rm eff}^2} \, k_8 \, . ~~
\end{eqnarray}
In addition we obtain expressions for $k_{15}^{(0)}$ and $k_{16}^{(0)}$ which are used below in the next paragraph.

Proceeding as above but taking terms up to first order in $\eta$ into account we find that the first order contributions of the $x$ operator expectation values $n_1$, $k_1$, and $k_2$ in Eq.~(\ref{coherences}) evolve according to
\begin{eqnarray} \label{48882}
\dot n_1^{(1)} &=& g_{\rm eff} \, k_2^{(1)} - \kappa \, n_1^{(1)}  \, , \nonumber \\
\dot k_1^{(1)} &=& - \delta_{\rm eff} \, k_2^{(1)} - \eta \nu \, k_{15}^{(0)} - {1 \over 2} \kappa \, k_1^{(1)} \, , \nonumber \\ 
\dot k_2^{(1)} &=& \delta_{\rm eff} \, k_1^{(1)} - \eta \nu \, k_{16}^{(0)} -  {1 \over 2} \kappa \, k_2^{(1)} \, . 
\end{eqnarray}
These equations form a closed set of cooling equations, when the above mentioned results for $k_{15}^{(0)}$ and $k_{16}^{(0)}$ are taken into account. Eliminating $n_1$, $k_1$ and $k_2$ adiabatically and neglecting all terms proportional to $\nu^2$, we find that
\begin{eqnarray} \label{zeroth2}
n_1^{(1)} &=& {32 \eta \nu \delta_{\rm eff} g_{\rm eff}^2 \over (\kappa^2 + 4 \delta_{\rm eff}^2)^2} \, k_8 \, . ~~
\end{eqnarray}
This means, $n_1^{(1)}$ follows the time evolution of $k_8$ adiabatically. 

In order to calculate $k_{11}^{(1)}$ and $k_{12}^{(1)}$, we need a closed set of cooling equations which applies up to first order in $\eta$ correctly. Applying Eq.~(\ref{dotA0}) again to $k_{11}$ and $k_{12}$ and $k_{15}$ to $k_{18}$, we find that the time derivatives of their first order corrections in $\eta$ are given by
\begin{eqnarray} \label{777}
\dot k_{11}^{(1)} &=& g_{\rm eff} \, k_{18}^{(1)} - \nu \, k_{12}^{(1)} + 2 \eta \nu \, n_3^{(0)} - \kappa \, k_{11}^{(1)} \, , \nonumber \\ 
\dot k_{12}^{(1)} &=& - g_{\rm eff} \, k_{15}^{(1)} + \nu \, k_{11}^{(1)} + 2 \eta \kappa \left[ n_1^{(0)} - n_3^{(0)} \right] - \kappa \, k_{12}^{(1)} \, , \nonumber \\
\dot k_{15}^{(1)} &=& - \delta_{\rm eff} \, k_{16}^{(1)} - \nu \, k_{18}^{(1)} + \eta \nu \left[ k_1^{(0)} + 2 k_{13}^{(0)} - k_{21}^{(0)} \right] \nonumber \\
&& + 2 \eta \kappa \, k_6^{(0)} - {1 \over 2} \kappa \, k_{15}^{(1)} \, , \nonumber \\
\dot k_{16}^{(1)} &=& \delta_{\rm eff} \, k_{15}^{(1)} + \nu \, k_{17}^{(1)} + \eta \nu \left[ k_2^{(0)} + 2 k_{14}^{(0)} - k_{22}^{(0)} \right] \nonumber \\
&& - 2 \eta \kappa \, k_5^{(0)} - {1 \over 2} \kappa \, k_{16}^{(1)}  \, , \nonumber \\
\dot k_{17}^{(1)} &=& - \delta_{\rm eff} \, k_{18}^{(1)} - \nu \, k_{16}^{(1)} + \eta \nu \left[ k_1^{(0)} + 2 k_5^{(0)} - k_{19}^{(0)} \right] \notag \\
&& - {1 \over 2} \kappa \, k_{17}^{(1)}  \, , \nonumber \\
\dot k_{18}^{(1)} &=& \delta_{\rm eff} \, k_{17}^{(1)} + \nu \, k_{15}^{(1)} + \eta \nu \left[ k_2^{(0)} + 2 k_6^{(0)} - k_{20}^{(0)} \right] \nonumber \\
&& - {1 \over 2} \kappa \, k_{18}^{(1)} \, .
\end{eqnarray}
Substituting the definitions of the mixed-particle expectation values $k_{13}$ and $k_{14}$ and $k_{19}$ to $k_{22}$ into Eq.~(\ref{dotA0}) and setting $\eta = 0$, we moreover find that 
\begin{eqnarray} \label{48222}
\dot k_{13} &=& - \delta_{\rm eff} \, k_{14} - {1 \over 2} \kappa \, k_{13} \, , \nonumber \\ 
\dot k_{14} &=& 2 g_{\rm eff} \, n_2 + \delta_{\rm eff} \, k_{13} - {1 \over 2} \kappa \, k_{14} \, , ~~ \nonumber \\
\dot k_{19} &=& - 2 g_{\rm eff} \, k_{10}  - \delta_{\rm eff} \, k_{20} - 2 \nu \, k_{22}  -  {1 \over 2} \kappa \, k_{19} \, , \nonumber \\ 
\dot k_{20} &=& \delta_{\rm eff} \, k_{19} + 2 \nu \, k_{21}  -  {1 \over 2} \kappa \, k_{20} \, , \nonumber \\
\dot k_{21} &=& - \delta_{\rm eff} \, k_{22} - 2 \nu \, k_{20} -  {1 \over 2} \kappa \, k_{21} \, , \nonumber \\ 
\dot k_{22} &=& 2 g_{\rm eff} \, k_9 + \delta_{\rm eff} \, k_{21} + 2 \nu \, k_{19} -  {1 \over 2} \kappa \, k_{22} \, .
\end{eqnarray}
These final six differential equations hold in zeroth order in $\eta$. Setting the right hand side of these and of the cooling equations in Eq.~(\ref{777}) equal to zero, we finally obtain the expressions
\begin{eqnarray} \label{k11121}
k_{11}^{(1)} &=& {16 \eta \nu g_{\rm eff}^2 \over (\kappa^2 + 4 \delta_{\rm eff}^2)^2} \left[ 2 \delta_{\rm eff} \, k_{10} + \kappa \right] \nonumber \\
&& + {64 \eta \nu g_{\rm eff}^4 \over \kappa (\kappa^2 + 4 \delta_{\rm eff}^2)^4} \, \left[ 5 \kappa^4 - 16 \kappa^2 \delta_{\rm eff}^2 - 16 \delta_{\rm eff}^4 \right] \, , \nonumber \\
k_{12}^{(1)} &=& {32 \eta \nu \delta_{\rm eff} g_{\rm eff}^2 \over (\kappa^2 + 4 \delta_{\rm eff}^2)^2} \,  \left[ 2 n_2 - k_9 + 1 \right] \nonumber \\
&& - {32 \eta g_{\rm eff}^4 (3 \kappa^2 - 4 \delta_{\rm eff}^2) \over (\kappa^2 + 4 \delta_{\rm eff}^2)^3} \, . 
\end{eqnarray}
Again we neglected terms proportional to $\nu^2$, since these are in general much smaller than the remaining terms.

\section{$n_1$, $k_{11}$, and $k_{12}$ in the strong confinement regime} \label{appC}

Let us now have a closer look at the strong confinement regime (cf.~Eq.~(\ref{cond})). However, different from the previous subsection, we no longer assume that some system parameters are much smaller than others. The reason that we nevertheless obtain relatively simple expressions for the quasi-stationary state solutions for $n_1$, $k_{11}$, and $k_{12}$ is that we eliminate in the following not only the $x$ and the mixed operator expectation values, but also the $y$ operator coherences $k_7$ to $k_{10}$. From Eq.~(\ref{4888}) we see that calculating $\dot n_2$ up to second order in $\eta$ requires knowing $n_1$ in zeroth order in $\eta$. Having a closer look at the above cooling equations, we see that the expression for $n_1^{(0)}$ in the strong confinement regime is the same as the expression in Eq.~(\ref{zeroth}). In addition, we need to calculate $k_{11}$ and $k_{12}$ up to first order in $\eta$.

Using again Eq.~(\ref{4888}), setting $\eta = 0$ and eliminating the $y$ operator coherences adiabatically from the system dynamics, we find that $k_7$ to $k_{10}$ all equal zero in zeroth order in $\eta$,
\begin{eqnarray} \label{nice}
k_7^{(0)} = k_8^{(0)} = k_9^{(0)} = k_{10}^{(0)} = 0 \, .
\end{eqnarray}
Taking this into account when eliminating the mixed operator expectation values $k_{11}$, $k_{12}$, and $k_{15}$ to $k_{18}$ in Eq.~(\ref{7772}) adiabatically, we now find that all of them vanish in zeroth order in $\eta$, 
\begin{eqnarray} \label{xxx}
k_{11}^{(0)} = k_{12}^{(0)} = 0 \, .
\end{eqnarray}
This means, the time derivative of $n_2$ in Eq.~(\ref{4888}) scales as $\eta^2$, at least to a very good approximation.

To calculate $k_{11}$ and $k_{12}$ up to first order in $\eta$, we have again a closer look at Eq.~(\ref{4888}). Using this equation and Eq.~(\ref{zeroth}), one can show that the $y$-coherences $k_7$ and $k_8$ are in first order in $\eta$ given by
\begin{eqnarray} \label{nice2}
k_7^{(1)} = {8 \eta \kappa g_{\rm eff}^2 \over \nu (\kappa^2 + 4 \delta_{\rm eff}^2)} \, , ~~ k_8^{(1)} =  {8 \eta g_{\rm eff}^2 \over \kappa^2 + 4 \delta_{\rm eff}^2}  \, .
\end{eqnarray}
Using Eqs.~(\ref{48222}) and (\ref{nice}), we see in addition that
\begin{eqnarray} \label{nice3}
&& k_{13}^{(0)} = - {8 \delta_{\rm eff} g_{\rm eff} \over \kappa^2 + 4 \delta_{\rm eff}^2} \, n_2 \, , ~~
k_{14}^{(0)} = {4 \kappa g_{\rm eff} \over \kappa^2 + 4 \delta_{\rm eff}^2} \, n_2 \, , \notag \\
&& k_{19}^{(0)} = k_{20}^{(0)} = k_{21}^{(0)} = k_{22}^{(0)} = 0 \, .
\end{eqnarray}
Applying Eq.~(\ref{dotA0}) again to $k_{11}$, $k_{12}$, and $k_{15}$ to $k_{18}$, we find that the time derivatives of the $k_{15}$ to $k_{18}$ in first order corrections in $\eta$ are now given by
\begin{eqnarray} \label{777z}
\dot k_{15}^{(1)} &=& - 2 g_{\rm eff} \, k_8^{(1)} - \delta_{\rm eff} \, k_{16}^{(1)} - \nu \, k_{18}^{(1)} \nonumber \\
&& + \eta \nu \left[ k_1^{(0)} + 2 k_{13}^{(0)} - k_{21}^{(0)} \right] + 2 \eta \kappa \, k_6^{(0)} - {1 \over 2} \kappa \, k_{15}^{(1)} \, , \nonumber \\
\dot k_{18}^{(1)} &=& 2 g_{\rm eff} \, k_7^{(1)} + \delta_{\rm eff} \, k_{17}^{(1)} + \nu \, k_{15}^{(1)} + \eta \nu \left[ k_2^{(0)} + 2 k_6^{(0)} \right.  \nonumber \\
&& \left. - k_{20}^{(0)} \right] - {1 \over 2} \kappa \, k_{18}^{(1)} \, .
\end{eqnarray}
while $k_{11}^{(1)}$, $k_{12}^{(1)}$, $k_{16}^{(1)}$, and $k_{17}^{(1)}$ evolve as stated in Eq.~(\ref{777}). Substituting Eqs.~(\ref{nice2}) and (\ref{nice3}) into these equations, using the solutions for $n_1^{(0)}$, $n_3^{(0)}$, and the coherences $k_1^{(0)}$, $k_2^{(0)}$, $k_5^{(0)}$ and $k_6^{(0)}$ which we obtained in App.~\ref{appB}, and eliminating $k_{11}^{(1)} $, $k_{12}^{(1)} $, and $k_{15}^{(1)} $ to $k_{18}^{(1)} $ adiabatically from the system dynamics, we obtain
\begin{eqnarray} \label{k111}
k_{11}^{(1)} &=& - {256 \eta \kappa \nu^2 \delta_{\rm eff} g_{\rm eff}^2 \over (\kappa^2 + 4 \delta_{\rm eff}^2) \mu^4} \, n_2 + {32 \eta \kappa g_{\rm eff}^4 \over \nu (\kappa^2 + 4 \delta_{\rm eff}^2)^2} \notag \\
&& + {16 \eta \kappa \nu g_{\rm eff}^2 \over (\kappa^2 + 4 \delta_{\rm eff}^2) \left[ \kappa^2 + 4 (\delta_{\rm eff} + \nu)^2 \right]} \, , \nonumber \\
k_{12}^{(1)} &=& {64 \eta \nu \delta_{\rm eff} g_{\rm eff}^2 \over (\kappa^2 + 4 \delta_{\rm eff}^2) \mu^4} \left[ \kappa^2 + 4 \delta_{\rm eff}^2 - 4 \nu^2 \right] n_2 \nonumber \\
&& + {32 \eta \nu g_{\rm eff}^2 (\delta_{\rm eff} + \nu) \over (\kappa^2 + 4 \delta_{\rm eff}^2) \left[ \kappa^2 + 4 (\delta_{\rm eff} + \nu)^2 \right]} \nonumber \\
&& + {32 \eta g_{\rm eff}^4 \over (\kappa^2 + 4 \delta_{\rm eff}^2)^2} 
\end{eqnarray}
with the constant $\mu^4$ defined as
\begin{eqnarray} \label{42222}
\mu^4 &\equiv& \left[\kappa^2 + 4 (\delta_{\rm eff} + \nu)^2 \right] \left[\kappa^2 + 4 (\delta_{\rm eff} - \nu)^2 \right] \, . ~~~~
\end{eqnarray}

\end{document}